%% file: main.tex
\documentclass[sigconf,natbib=true]{acmart}

\usepackage{acronym}
\input{acronyms}

\input{temp-authors-commands}
\AtBeginDocument{%
  }

\setcopyright{acmlicensed}
\copyrightyear{2018}
\acmYear{2018}
\acmDOI{XXXXXXX.XXXXXXX}

\acmConference[SIGIR '25]{Make sure to enter the correct
  conference title from your rights confirmation email}{June 03--05,
  2018}{Woodstock, NY}
\acmISBN{978-1-4503-XXXX-X/18/06}

\usepackage{algorithmic}
\usepackage{adjustbox}
\usepackage{subfig}

\newcommand{\partitle}[1]{\vspace{1mm}\noindent\textbf{#1.}}

\begin{document}

\title{Judging the Judges: \\ A Collection of LLM-Generated Relevance Judgements}


\author{
Hossein A.~Rahmani
}
\orcid{0000-0002-2779-4942} 
\affiliation{%
        \institution{University College London}
        \city{London}
        \country{UK}
}
\email{hossein.rahmani.22@ucl.ac.uk}

\author{Clemencia Siro}
\orcid{0000-0001-5301-4244} 
\affiliation{%
        \institution{University of Amsterdam}
        \city{Amsterdam}
        \country{The Netherlands}
}
\email{c.n.siro@uva.nl}

\author{
Mohammad Aliannejadi
}
\orcid{0000-0002-9447-4172} 
\affiliation{%
        \institution{University of Amsterdam}
        \city{Amsterdam}
        \country{The Netherlands}
}
\email{m.aliannejadi@uva.nl}

\author{Nick Craswell}
\orcid{0000-0002-9351-8137} 
\affiliation{%
        \institution{Microsoft}
        \city{Bellevue}
        \country{US}
}
\email{nickcr@microsoft.com}

\author{Charles L.~A.~Clarke}
\orcid{0000-0001-8178-9194}
\affiliation{%
  \institution{University of Waterloo}
  \city{Waterloo, Ontario}
  \country{Canada}
}
\email{claclark@gmail.com}

\author{Guglielmo Faggioli}
\orcid{0000-0002-5070-2049} 
\affiliation{%
        \institution{University of Padua}
        \city{Padua}
        \country{Italy}
}
\email{faggioli@dei.unipd.it}

\author{Bhaskar Mitra}
\orcid{0000-0002-5270-5550} 
\affiliation{%
        \institution{Microsoft}
        \city{Montréal}
        \country{Canada}
}
\email{bmitra@microsoft.com}

\author{Paul Thomas}
\orcid{0000-0003-2425-3136} 
\affiliation{%
        \institution{Microsoft}
        \city{Adelaide}
        \country{Australia}
}
\email{pathom@microsoft.com}

\author{Emine Yilmaz}
\orcid{0000-0003-4734-4532} 
\affiliation{%
        \institution{University College London \& Amazon}
        \city{London}
        \country{UK}
}
\email{emine.yilmaz@ucl.ac.uk}
\renewcommand{\shortauthors}{Rahmani et al.}

\begin{abstract}
Using Large Language Models (LLMs) for relevance assessments offers promising opportunities to improve Information Retrieval (IR), Natural Language Processing (NLP), and related fields. Indeed, LLMs hold the promise of allowing IR experimenters to build evaluation collections with a fraction of the manual human labor currently required. This could help with fresh topics on which there is still limited knowledge and could mitigate the challenges of evaluating ranking systems in low-resource scenarios, where it is challenging to find human annotators. Given the fast-paced recent developments in the domain, many questions concerning LLMs as assessors are yet to be answered. Among the aspects that require further investigation, we can list the impact of various components in a relevance judgment generation pipeline, such as the prompt used or the LLM chosen.

This paper benchmarks and reports on the results of a large-scale automatic relevance judgment evaluation, the \textit{LLMJudge challenge} at SIGIR 2024, where different relevance assessment approaches were proposed. In detail, we release and benchmark 42 LLM-generated labels of the TREC 2023 Deep Learning track relevance judgments produced by eight international teams who participated in the challenge. Given their diverse nature, these automatically generated relevance judgments can help the community not only investigate systematic biases caused by LLMs but also explore the effectiveness of ensemble models, analyze the trade-offs between different models and human assessors, and advance methodologies for improving automated evaluation techniques.
The released resource is available at the following link: \url{https://llm4eval.github.io/LLMJudge-benchmark/}. 
\end{abstract}





\maketitle

\input{sections/01-introduction}
\input{sections/02-relatedwork}
\input{sections/03-llmjudge}
\input{sections/04-submissions}
\input{sections/05-results}
\input{sections/06-conclusion}

\begin{acks}
The challenge is organized as a joint effort by the University College London, Microsoft, the University of Amsterdam, the University of Waterloo, and the University of Padua. The views expressed in the content are solely those of the authors and do not necessarily reflect the views or endorsements of their employers and/or sponsors. This research is supported by the Engineering and Physical Sciences Research Council [EP/S021566/1], CAMEO, PRIN 2022 n.~2022ZLL7MW and by the Dreams Lab, a collaboration between Huawei Finland, the University of Amsterdam, and the Vrije Universiteit Amsterdam.
\end{acks}


\bibliographystyle{ACM-Reference-Format}
\bibliography{references}

\clearpage

\appendix


\input{tables/tbl_labels}

\input{tables/tbl_alpha}

\end{document}

%% file: acronyms.tex
\acrodef{IR}[IR]{Information Retrieval}
\acrodef{LLM}[LLM]{Large Language Model}

%% file: sections/01-introduction.tex
\section{Introduction}
\label{sec:introduction}
The Cranfield paradigm has been the standard \ac{IR} evaluation methodology since the 1960s \cite{Cleverdon1960TheAC,10.5555/275537.275544}. This methodology requires three components to evaluate an \ac{IR} system: a document corpus, topics (queries), and \textit{corresponding relevance judgments} that indicate which documents are relevant to each topic.

In practice, an \ac{IR} system processes the topics to retrieve relevant documents from the corpus, while the relevance judgments serve as the ground truth for measuring system effectiveness. Of these three components, the first two are relatively straightforward to acquire. The document corpus can be constructed to mirror the target domain of the \ac{IR} system through various collection methods, such as web crawling, newspaper archives, or scientific literature databases. Similarly, topics can be manually handcrafted by non-experts or through the systematic analysis of query logs \cite{DBLP:conf/nips/NguyenRSGTMD16}, ensuring they represent an appropriate sample of the expected query space.


The main challenge lies in obtaining relevance judgments. These judgments map topics to documents, specifically indicating how well each document satisfies the information need expressed in the topic. Creating complete relevance judgments requires significant human effort and careful quality control to ensure consistency and reliability~\citep{DBLP:journals/ftir/Sanderson10}.

Over time, three major approaches became the de facto standard to collect relevance judgments. The first approach, followed by the major evaluation campaigns such as TREC~\cite{DBLP:journals/ipm/Jones95}, NTCIR~\cite{kando1999overview} and CLEF~\cite{DBLP:conf/clef/2000}, is based on editorial assessments. In this case,  professional assessors judge whether a document is relevant in response to the topic. While these judgments are often of very high quality, they are also expensive to obtain in terms of time and cost~\cite{DBLP:journals/ftir/Sanderson10}.
A second strategy is based on employing crowd-assessors to annotate the documents. While crowd annotations are typically less expensive than editorial annotations, they are also qualitatively inferior. Crowd annotations often contain much more noise and errors.
Thirdly, annotations can be obtained as implicit feedback from user - \ac{IR} system interaction. These annotations are virtually free as they are based mostly on already available data (e.g., click logs) and can embed user-specific characteristics, such as their knowledge, tastes, and personal inclinations. Nevertheless, implicit feedback is also affected by noise, biases~\cite{JoachimsSwaminathanSchnabel2017}, and privacy problems \cite{DBLP:journals/tissec/AonghusaL16}.
In general, the types of annotations can be organized in a spectrum: on one side we have accurate and costly editorial judgments, in the middle we have labels produced by the crowd, on the other side, we find inexpensive but imprecise and biased implicit feedback.

\acp{LLM} have recently emerged as a promising fourth approach to gather relevance judgments~\cite{DBLP:conf/sigir/MacAvaneyS23,faggioli2023perspectives,thomas2023large,rahmani2024synthetic}. Initial experiments show that LLMs can achieve comparable performance to crowd workers on standard \ac{IR} tasks~\citep{DBLP:conf/sigir/BlancoHHMPTT11} and potentially reduce annotation costs. 
The \texttt{LLMJudge} challenge \cite{rahmani2024llmjudge} was organized as part of the \texttt{LLM4Eval}\footnote{\url{https://llm4eval.github.io/}} workshop~\cite{rahmani2024llm4eval} at SIGIR 2024 as a shared task to study the effectiveness of using \acp{LLM} as annotation tools.
While \acp{LLM} have shown to be effective annotation tools, several aspects are yet to be understood. For example, it is not clear the impact of changing the prompt, which biases are present in the \ac{LLM}-generated relevance judgments, and if there is a risk of evaluation circularity~\cite{faggioli2023perspectives}. Beyond these challenges, there is also a need to explore the effectiveness of ensemble models, examine the trade-offs between different LLM-based and human assessments, and develop more advanced methodologies to enhance automated evaluation techniques. We release the relevance annotations produced by the teams participating in the \texttt{LLMJudge} challenge, to help the community investigate these aspects linked to using \ac{LLM} as automatic annotation tools.



Our contributions are the following:
\begin{itemize}
    \item We release 42 pools of automatically generated relevance judgments produced by 8 different research teams that participated in the \texttt{LLMJudge} challenge.
    \item We confirm current observations about the state of the art, noticing that, while many approaches maintain ranking consistency, their absolute scoring tendencies differ, potentially introducing biases in evaluation.
    \item From the methodological perspective, we investigate several approaches that can be adopted to assess the effectiveness of an LLM-based relevance judgment process and provide a set of figures that will serve future researchers as baselines. 
\end{itemize}

The remainder of this paper is organized as follows: Section~\ref{sec:relatedwork} introduces the related works, including the challenges associated with automatically generated relevance judgments. Section~\ref{sec:llmjudge_resource} delineates the structure and describes the collection of the LLMJudge resource. Section \ref{sec:intro-methods} summaries submission runs of \texttt{LLMJudge} challenge. In Section~\ref{sec:results}, we analyze the dataset and provide some insights on the \ac{LLM}-generated judgments. Finally, in Section~\ref{sec:conclusion}, we draw our conclusion and outline our future work.

%% file: sections/02-relatedwork.tex
\section{Related Work}
\label{sec:relatedwork}
Traditionally, an experimental \ac{IR} collection includes three elements, a corpus, a set of topics, and the relevance judgments, defining which documents are relevant in response to the topics.
Over the last 30 years, since the first TREC campaign~\cite{DBLP:conf/trec/1992}, the most common strategy to obtain such relevance judgments has involved expert annotators, capable of providing the most accurate labels. 
The cost of this process can be partially reduced with pooling~\cite{croft2009search}, but the monetary and temporal costs of building an \ac{IR} experimental collection following this paradigm remain extremely high.

Automatic relevance judgment has recently received significant attention in the IR community. In earlier studies, ~\citet{faggioli2023perspectives} studied different levels of human and LLMs collaboration for automatic relevance judgment. They suggested the need for humans to support and collaborate with LLMs for a human-machine collaboration judgment. ~\citet{thomas2023large} leverage LLMs capabilities in judgment at scale, in Microsoft Bing. They used real searcher feedback to build an LLM and prompt in a way that matches the small sample of searcher preferences. Their experiments show that LLMs can be as good as human annotators in indicating the best systems. They also comprehensively investigated various prompts and prompt features for the task and revealed that LLM performance on judgments can vary with simple paraphrases of prompts. Recently, \citet{rahmani2024synthetic} have studied fully synthetic test collection using LLMs. In their study, they generated synthetic queries and synthetic judgment to build a full synthetic test collation for retrieval evaluation. They have shown that LLMs can generate a synthetic test collection that results in system ordering performance similar to evaluation results obtained using the real test collection.

On a different line, \citet{DBLP:conf/sigir/Dietz24} defines a LLM-based ``autograding'' approach. This evaluation strategy targets generated content that cannot be evaluated in a purely offline scenario and it consists of using a question bank as the evaluation test-bed. An \ac{LLM} measures the effectiveness of the generative model in answering the questions, possibly with the supervision of a human. The autograding approach proposed by \citet{DBLP:conf/sigir/Dietz24} includes an automatic passage evaluation whose task aligns with the one evaluated in \texttt{LLMJudge}.

\subsection{Criticisms and Open Challenges}
The use of \acp{LLM} as assessors comes with major bias risks and challenges that should not be neglected, especially considering the impact they might have in the development of \ac{IR} evaluation.

\partitle{Bias}
First and most importantly, \acp{LLM} are affected by bias~\cite{DBLP:conf/fat/BenderGMS21}. Their internal representation of the concepts is, by construction, conditioned on the context such concepts appear in~\cite{DBLP:conf/nips/VaswaniSPUJGKP17}. Thus, depending on the underlying data, the \ac{LLM} might form a biased notion of relevance that might reflect upon the relevance judgments generated by it. Quantifying the bias, identifying its source, and mitigating its consequences are still open issues that need to be addressed. We hope that the release of this collection will help the research community with the needed data to study how to deal with the bias in \ac{LLM}-generated relevance judgments.

\partitle{Circularity}
A second source of concern when it comes to using \acp{LLM} as assessors relates to the risk of \textit{circular evaluation}~\cite{faggioli2023perspectives,DBLP:journals/corr/abs-2409-15133}. For example, the same \ac{LLM} might be used to generate relevance judgments and as a document ranker. This would induce a strong bias on the validity and generalizability of the relevance judgments.

\partitle{Environmental Impact}
An often hidden cost of the \acp{LLM} concerns their environmental impact in terms of energy utilization, carbon emissions~\cite{DBLP:journals/corr/abs-2408-09713,DBLP:conf/sigir/ScellsZZ22}, and water consumption~\cite{DBLP:conf/ictir/ZucconSZ23}.
While \acp{LLM} might allow building collections at a fraction of the monetary and temporal cost, we should account for the environmental impact of such a process, limiting our reliance on ``disposable'' relevance judgments.

\partitle{Vulnerability to Attacks and Adversarial Misuse}
\citet{DBLP:conf/ecir/ParryFMPH24} and \citet{DBLP:conf/sigir-ap/Alaofi0SS24} illustrate the vulnerability of the \acp{LLM} to mischievous manipulations of the corpus. For example,~\citet{DBLP:conf/ecir/ParryFMPH24} show that, by introducing keywords such as the term ``relevant'' in a document, it will more likely considered relevant by an \ac{LLM}. Similar behavior is observed also by \citet{DBLP:conf/sigir-ap/Alaofi0SS24}, who notice that by introducing the query on the document, more probably an \ac{LLM} will consider the document relevant to such a query --- even if the rest of the document is composed by random terms.
More recently, \citet{DBLP:journals/corr/abs-2412-17156} show how, by properly crafting an adversarial run, it is possible to cheat an \ac{LLM} used as an assessor. \citet{DBLP:journals/corr/abs-2412-17156} crafted a run following the same approach used by~\citet{upadhyay2024umbrela} to pool the documents and build the \ac{LLM}-generated relevance judgments used for TREC 2024 RAG. Such a run achieved consistently higher effectiveness under the fully automatic evaluation paradigm compared to its performance based on manual relevance judgments. 

By releasing this collection of \ac{LLM}-generated relevance judgments we want to foster the analysis and study of possible sources of biases and systematic errors, to mitigate them and allow for the development of more effective and robust future solutions that involve \acp{LLM} as tools to support the annotation process.

%% file: sections/03-llmjudge.tex
\section{LLMJudge Resource}
\label{sec:llmjudge_resource}
This section details how we designed the \texttt{LLMJudge} challenge task, the data construction process, and the evaluation metrics.

\begin{figure}
    \centering
    \subfloat[Dev set\label{fig:sample-dev}]
    {
        {
            \includegraphics[scale=0.26]{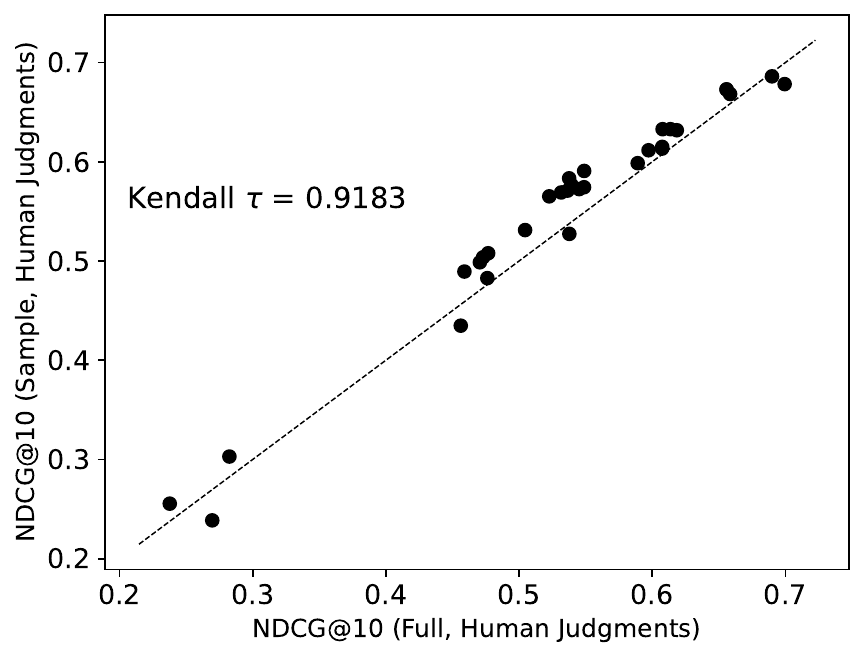}
        }
    }%
    \subfloat[Test set\label{fig:sample-test}]
    {
        {
            \includegraphics[scale=0.26]{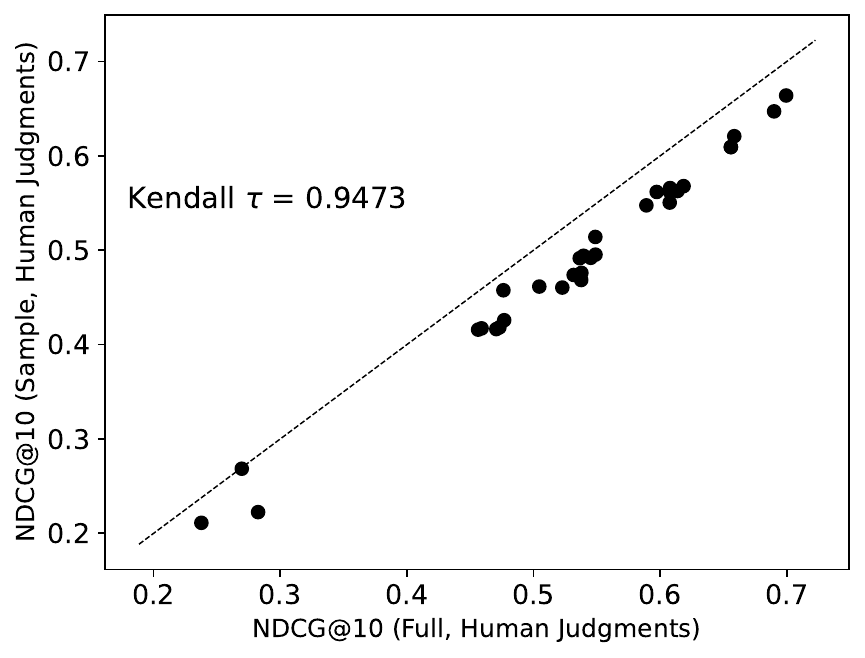}
        }
    }%
    \caption{Samples correlation with TREC 2023 DL full qrel}%
    \label{fig:data-samples}%
\end{figure}

\subsection{LLMJudge Task}
The task of the LLMJudge challenge is, given the query and passage as input, how they are relevant. Similar to TREC 2023 Deep Learning track \cite{craswell2024overview}, we use \textit{four-point scale} judgments to evaluate the relevance of the query to the passage as follows:

\begin{itemize}
    \item\textbf{[3] Perfectly relevant}: The passage is dedicated to the query and contains the exact answer. 
    \item\textbf{[2] Highly relevant}: The passage has some answers for the query, but the answer may be a bit unclear, or hidden amongst extraneous information. 
    \item\textbf{[1] Related}: The passage seems related to the query but does not answer it. 
    \item\textbf{[0] Irrelevant}: The passage has nothing to do with the query. 
\end{itemize}

More specifically, the LLMJudge challenge is, by providing the datasets that include queries, passages, and query-passage files to participants, to ask LLMs to generate a score [0, 1, 2, 3] indicating the relevance of the query to the passage.

\subsection{LLMJudge Data}
The \texttt{LLMJudge} challenge dataset is built upon the passage retrieval task dataset of the TREC 2023 Deep Learning track\footnote{\url{https://microsoft.github.io/msmarco/TREC-Deep-Learning.html}} \cite{craswell2024overview}. The TREC 2023 Deep Learning track qrel consists of 82 queries, including 51 real queries and 31 synthetic queries (13 generated by T5 and 18 generated by GPT-4). To create a dev and test set similar to the TREC 2023 Deep Learning track full qrel, we randomly sampled 15 queries from 51 real queries, 5 queries from T5 queries, and 5 queries from GPT-4 queries for each set. Figure \ref{fig:data-samples} shows Kendall's $\tau$ correlation of the TREC 2023 Deep Learning track run submission on LLMJudge sampled dev and test sets with the TREC 2023 Deep Learning track full qrel. Table \ref{tbl:llmjudge-dataset} shows the statistics of the \texttt{LLMJudge} challenge datasets. The test set is used for the generation of judgment by participants, while the development set could be used for few-shot or fine-tuning purposes.

\input{tables/tbl_dataset}

\subsection{Evaluation}
We evaluate submission results on two different levels, the correlation of the judgments and the ranking correlation of systems evaluated using judgment submissions:

\begin{itemize}
    \item \textbf{Label Correlation.} We use the automated evaluation metrics Cohen's Kappa ($\kappa$) and Krippendorff's Alpha ($\alpha$) on human judgments and the judgments submitted by participants;
    \item \textbf{System Ranking Correlation.} We use Kendall's Tau ($\tau$) and Spearman's rank ($\rho$) correlation to evaluate the system ordering of TREC 2023 Deep Learning Track \cite{craswell2024overview} submitted systems on human judgments and participants' LLM-based judgments.
\end{itemize}

We use \texttt{scikit-learn}\footnote{\url{https://scikit-learn.org/stable/index.html}} to compute Cohen's $\kappa$, Kendall's $\tau$, Spearman's $\rho$. Krippendorff's $\alpha$ is also calculated using the Fast Krippendorff\footnote{\url{https://github.com/pln-fing-udelar/fast-krippendorff}} Python package. 

\subsection{Publicly Available Resources}
To facilitate research in the area we have made the LLMJudge dataset, sample prompt, quick starter for automatic judgment, submitted runs, prompts, codes for quick starting the evaluation, and more detailed results publicly available on the \texttt{LLMJudge} webpage at: \url{https://llm4eval.github.io/LLMJudge-benchmark/}.

%% file: tables/tbl_dataset.tex
\begin{table}
    \caption{Statistics of LLMJudge Dataset}
    \label{tbl:llmjudge-dataset}
        \begin{tabular}{lcc}
            \toprule
            \textbf{} & \textbf{Dev} & \textbf{Test} \\
            \midrule
             \# queries  & 25    & 25 \\
             \# passage & 7,224 & 4,414 \\
             \# qrels    & 7,263 & 4,423 \\
             \midrule
             \# irrelevant (0)         & 4,538 & 2,005 \\
             \# related (1)            & 1,403 & 1,233 \\
             \# highly relevant (2)    & 625 & 808 \\
             \# perfectly relevant (3) & 697 & 377 \\
            \bottomrule
        \end{tabular}
\end{table}

%% file: sections/04-submissions.tex
\section{Submitted Runs}
\label{sec:intro-methods}
We provide all submitted runs as a resource for future research and comparison. The submissions include 9 \emph{baseline approaches} developed by the organizers and 33 \emph{methods from participating teams}. Analysis of these submissions reveals several methodological directions in \ac{LLM}-based relevance assessment, focusing on \emph{prompting techniques, model adaptation, multi-phase evaluation, aggregation strategies, and classification-based refinement}.  

\input{tables/tbl_info}

Most submissions implement either \emph{direct prompting} or \emph{criteria decomposition pipelines}. \emph{Direct prompting methods} range from simple \emph{relevance scoring instructions} to \emph{chain-of-thought reasoning}, where \acp{LLM} justify their judgments before assigning a score. Some approaches explore \emph{zero-shot prompting}, while others incorporate \emph{semantic label assignments, linguistic alignment, or multi-prompt aggregation} to improve consistency and reduce overestimation biases.  
Beyond prompting, some teams \emph{fine-tune LLMs on relevance datasets}, including \emph{TREC Deep Learning track qrels} and the \emph{LLMJudge development set}, testing different \emph{model sizes (8B vs.~70B)} to assess the impact of adaptation on evaluation performance.  
A subset of submissions structures evaluation into \emph{multi-phase pipelines}, applying \emph{binary filtering before graded scoring}, \emph{question-based reasoning}, or \emph{decision trees} to refine assessments. Other approaches decompose relevance into \emph{specific dimensions} such as \emph{exactness, coverage, topicality, and contextual fit}, or employ \emph{nugget-based assessments} for more granular judgments.  
To enhance robustness, several methods \emph{combine outputs from multiple prompts or models} using \emph{multi-prompt averaging, binary-to-graded conversions, or conservative ensembling} to stabilize scores. Others treat \emph{relevance assessment as a classification task}, extracting features from LLM outputs and training \emph{Machine Learning classifiers} to refine final scores and improve alignment with human judgments.  

Below we detail the baselines and a summary of the submitted runs. We also summarize the submission details in Table \ref{tab:infos}.

\partitle{LLMJudge Baseline}
The baseline judges provided by the LLMJudge challenge organizers serve as reference methods for evaluation. Three distinct approaches are proposed as baselines: \texttt{llmjudge\\-simple}, \texttt{llmjudge-cot}, and \texttt{llmjudge-thomas}. The \texttt{llmjudge-\\simple} method employs a straightforward prompt, instructing the model to directly provide a relevance judgment based on the query and passage. In contrast, \texttt{llmjudge-cot} adopts a chain-of-thought (CoT) approach, prompting the model to articulate its reasoning process before delivering a judgment. Lastly, \texttt{llmjudge-thomas} incorporates the prompt design introduced by \cite{thomas2023large}, offering an alternative strategy for evaluation.

\partitle{NISTRetrieval-instruct}
This is a submission from NIST which has three different variants, namely, \texttt{NISTRetrieval-instruct0}, \texttt{NISTRetrieval-instruct1}, and \texttt{NISTRetrieval-instruct2} that aims to investigate the reproducibility of the method proposed by \citet{thomas2023large} and the reproducibility capabilities of LLMs when we used them for automatic relevance judgment.

\partitle{NISTRetrieval-reason}
Similar to \texttt{NISTRetrieval-instruct}, this NIST submission includes three related methods -- \texttt{NISTRetrieval-\\reason0}, \texttt{NISTRetrieval-reason1}, and \texttt{NISTRetrieval-reason2}. The team observed that prompting LLMs to provide reasoning across various tasks could improve response quality. To examine whether this approach could also enhance relevance judgment, they modified the prompt from \citet{thomas2023large} to allow the LLM to generate reasoning. These three runs were included to assess the reproducibility capabilities of LLMs when used for evaluation.

\partitle{Prophet-setting}
This method builds on the idea of fine-tuning an LLM with different available datasets for automatic relevance judgment, as described in \cite{meng2024query}\footnote{Code is available at \url{https://github.com/ChuanMeng/QPP-GenRE}}. Specifically, they fine-tuned \texttt{Llama-3-8B} under three different settings, training the model for five epochs in each. These settings include: \texttt{Prophet-setting1}, fine-tuned on the LLMJudge development set; \texttt{Prophet-setting2}, fine-tuned on the qrels of TREC-DL 2019, 2020, and 2021; and \texttt{Prophet-setting4}, which combines fine-tuning on the qrels of TREC-DL 2019, 2020, and 2021 with the LLMJudge development set.

\partitle{William-umbrela1}
This approach is zero-shot prompting the LLM to produce relevance assessments. They used UMBRELA \cite{upadhyay2024umbrela} to generate relevance judgments using the prompting technique suggested by \citet{thomas2023large}. The team mentioned that ``\textit{I tried many different approaches, but I did not manage to find anything that really seemed to consistently improve on zero-shot. It seemed like this dataset may have been harder and/or noisier than others referenced in the literature -- on my development set it was hard to get $> 0.3$ Cohen's $\kappa$, whereas the literature mentions values of $0.4$ up to $0.6$ even.}''.

\partitle{William-umbrela2}
The main idea of this method is to take the approach from the UMBRELA \cite{upadhyay2024umbrela} -- zero-shot prompting technique from \citet{thomas2023large}, but to see if the performance would be improved by asking the model to output semantic labels (i.e., \textit{Irrelevant}, \textit{Related}, \textit{Highly relevant}, \textit{Perfectly relevant}), rather than a numerical score (i.e., 0, 1, 2, 3).

\partitle{William-umbrela3}
This method is an ensemble of~ \texttt{William-\\umbrela1} and \texttt{William-umbrela2} approaches by taking the \textit{min}. The team mentioned that ``\textit{The logic behind using min as an aggregator is that in this dataset, it pays to be conservative in the rating. They also said that on a subset of the training data that they held out for testing, this ensembling approach outperformed either of the two other approaches (i.e., William-umbrela1 and William-umbrela2)''}.

\partitle{H2oloo-fewself}
This method uses the best prompt proposed by Thomas et al.~\cite{thomas2023large} to instruct GPT-4o. It incorporates few-shot examples to guide the model in distinguishing between relevant labels effectively.

\partitle{H2oloo-zeroshot1}
This method fined-tuned a Llama-8B using the TREC DL 2019 to 2022 qrels for relevance judgment prediction.

\partitle{H2oloo-zeroshot2}
This method fined-tuned a Llama-8B using the TREC DL 2019 to 2022 qrels and the LLMJudge dev set qrel for relevance judgment prediction.

\partitle{Olz-gpt4o}
This method uses a simple prompt where they just ask for the relevance judgment without any special techniques. The idea is to see how models can solve relevance judgment tasks without considering any particular prompting or fine-tuning techniques. The primary goal is to assess whether a low-effort prompt could reliably derive relevance labels from LLMs that are practically usable.

\partitle{Olz-exp}
This method is similar to \texttt{Olz-gpt4o} but they also asked LLM to reason its judgment as part of the evaluation.

\partitle{Olz-halfbin}
This method leverages Llama-3 models with $8B$ and $70B$ parameters to assess document relevance using nine distinct prompts. These prompts are divided into two categories: four \emph{graded relevance prompts}, which instruct the model to assign a score from 0 to 3 with slight instruction variations, and five \emph{binary relevance prompts}, which require binary judgments with different definitions of relevance.  
Both model variants generate outputs for all nine prompts. These outputs serve as features for training a logistic regression classifier, which produces the final graded labels. Training is conducted using labels generated by GPT-4o (via the \texttt{Olz-gpt4o} method) rather than the standard development set annotations, based on the assumption that the development and test set labels may have been derived using different methods. Analyzing these discrepancies, the team found GPT-4o’s judgments more aligned with their expectations, leading to its adoption as the primary reference for training.

\partitle{Olz-somebin}
The procedure of this method is identical to the \texttt{Olz-halfbin} method, except the logistic regression classifier was trained on the provided development set labels instead of those generated by GPT-4o (using \texttt{Olz-gpt4o} method).

\partitle{Olz-multiprompt}
This method, instead of using a classifier like \texttt{Olz-halfbin} and \texttt{Olz-somebin}, directly aggregated the relevance judgments by averaging. The binary labels were first scaled by multiplying them by three (to convert them into $0$ or $3$). Then a simple average was calculated across the nine prompts and rounded on a scale of 0 to 3, and the resulting value served as the final graded label.

\partitle{RMIT-IR}
This submission introduces three relevance assessors, \texttt{RMITIR-GPT4o}, \texttt{RMITIR-llama38b}, and \texttt{RMITIR-llama70B}. The proposed approach begins by having the LLM provide a binary relevance judgment to filter out irrelevant queries and improve irrelevance filtering. Next, three scores are generated, and averaged, and the result is rounded to produce the final score. The method was tested using three different LLMs: GPT-4o (\texttt{RMITIR-GPT4o}), Llama3-8B (\texttt{RMITIR-llama38b}), and Llama3-70B (\texttt{RMITIR-llama70B}). The team noted that ``\textit{GPT-4o appears to be the best-performing model based on our experiences}.''

\partitle{TREMA-4prompt}   
This method evaluates passage relevance by decomposing it into four specific criteria: exactness (how precisely the passage answers the query), coverage (proportion of content discussing the query), topicality (subject alignment between passage and query), and contextual fit (presence of relevant background). The evaluation follows a two-phase process where each criterion is first assessed independently and then combined through a final prompt to determine the overall relevance label. Full details of the criteria and rationale are provided in \cite{farzi2024best}.

\partitle{TREMA-CoT}
This method implements a chain-of-thought evaluation process inspired by \citet{Sun2023IsCG}. The approach consists of three phases: First, the \ac{LLM} makes a binary relevance judgment (yes/no) of the passage. Based on this judgment, different relevance criteria are evaluated in the second phase - for relevant passages, exactness and coverage are assessed, while non-relevant passages are evaluated on contextual fit and topicality (all scored 0-3). In the final phase, these scores determine the overall relevance label: relevant passages receive labels 2-3 based on exactness and coverage scores, while non-relevant passages receive labels 0-1 based on contextual fit and topicality assessment.

\partitle{TREMA-other}
This approach investigates whether aligning the linguistic styles of queries and passages can enhance relevance judgments. In the first phase, the LLM generates a query-like representation for each passage, designed to match the query's linguistic style and length. This generated query serves as a summary of the passage's content, formatted in a way that aligns with typical query phrasing. In the second phase, the LLM evaluates the similarity between the original query and the generated query on a scale from 0 to 3, corresponding to the relevance labeling system. Higher similarity scores indicate a stronger alignment between the passage's content and the query's intent. This method integrates linguistic style alignment with content relevance to improve relevance labeling.

\partitle{TREMA-sumdecompose}
This method consists of two phases. Phase one is identical to the \texttt{TREMA-4prompt} method, where the ``relevance'' is decomposed into four criteria, leading to four criteria-specific grades.  In Phase Two, the individual grades from Phase One are summed to produce a total grade. Based on this total, a final relevance label between 0 and 3 is assigned to each query-passage pair: a total grade of 10-12 yields a relevance label of 3, 7-9 yields a relevance label of 2, 5-6 yields a relevance label of 1, and 0-4 yields a relevance label of 0.

\partitle{TREMA-naiveBdecompose}
This method consists of two phases. Phase one is identical to the \texttt{TREMA-4prompt} method, where the ``relevance'' is decomposed into four criteria, leading to four criteria-specific grades. In phase two, these decomposed grades are aggregated into a final relevance label using a Gaussian Naive Bayes model, implemented with Scikit-learn's GaussianNB() classifier. The model is trained on the decomposed feature grades and then predicts the relevance label for each passage.

\partitle{TREMA-rubric0}
This method is based on the RUBRIC Autograder Workbench \cite{dietz2024workbench}. This method defines the relevance of the query via 10 open-ended questions. The questions are generated using the ChatGPT 3.5 model. Each passage is scanned whether it is possible to answer each of the questions (and how well), which is captured as a grade. They use the FLAN-T5-large LLM from Huggingface to grade the answerability from 0 (worst) to 5 (best). Details and prompts are available in the Workbench benchmark \cite{dietz2024workbench}. The grades are mapped to relevance labels by a heuristic mapping on the second-highest grade achieved on any of the questions. Grade 5 is mapped to relevance label 3, grade 4 is mapped to label 1 and all other grades are mapped to label 0. This was the best manual mapping on the dev set \cite{farzi2024rubric}.

\partitle{TREMA-questions}
Same question and grading as in \texttt{TREMA\\-rubric0}, but uses a more elaborate calibration for converting grades to relevance labels, based on \texttt{scikit-learn}'s ExtraTrees classifier. The classifier is based on features that include ranked grades for each question (sorted in descending order), ranked question difficulty (based on average grades across the pool), and counts of correct answers at various grade thresholds (e.g., number of answers graded 5, 4 or better, etc.). Each of these features is encoded using both one-hot and numerical representations to capture detailed information about question-based relevance. The classifier is trained on the dev set.

\input{tables/tbl_results}

\partitle{TREMA-nuggets}
Same approach as \texttt{TREMA-questions}, but uses 10 open-ended key fact nuggets instead of questions, along with an adapted prompt that assesses whether key facts are mentioned in the passage. The same ExtraTrees classifier with the same features is used for converting grades into relevance labels.

\partitle{TREMA-direct}
This approach focuses exclusively on features of direct relevance
labeling methods, which instruct an LLM to judge whether a passage is relevant for a query, using a variety of prompts from \citet{Sun2023IsCG}, \citet{faggioli2023perspectives}, and HELM \cite{liang2022holistic}. The model excludes question-based and nugget-based features, simplifying its input to focus solely on the predictive power of direct labeling. The relevance labels are obtained with an ExtraTrees classifier trained on the dev set. Features include binary or multi-class predictions from labeling approaches. Each label is encoded using both one-hot and numerical encodings to capture both categorical and ordinal aspects of the predictions. This approach is computationally lighter than TREMA-all and serves as a baseline to evaluate how well direct relevance labels alone can predict passage relevance.

\partitle{TREMA-all}
This approach incorporates all features from \texttt{TREMA-\\questions}, \texttt{TREMA-nuggets}, and \texttt{TREMA-direct} approaches via a single ExtraTrees classifier that is trained on the dev set.

%% file: tables/tbl_info.tex
\begin{table}
    \centering
    \caption{LLMJudge challenge submissions details. Ensemble (Ens.) indicates if submissions combine multiple judges or use them as features to train a classifier for judgment. LR: Logistic Regression, ET: ExtraTrees, GaussianNB: Gaussian Naive Bayes are classifiers. If a submission used multiple prompts, we consider the more advanced one (CoT > Zero-Shot) in this table. FT: Fine-Tuning, N: Numerical, S: Semantic.}
    \adjustbox{max width=\columnwidth}{%
    \begin{tabular}{llccccc}
    \toprule
    \textbf{Submission ID} & \textbf{Model} & \textbf{Size} & \textbf{FT} & \textbf{Prompt} & \textbf{Label} & \textbf{Ens.} \\
    \midrule
    NISTRetrieval-instruct0 & Llama-3-Instruct & 8B & - & Zero-shot & N & - \\
    NISTRetrieval-instruct1 & Llama-3-Instruct & 8B & - & Zero-shot & N & - \\
    NISTRetrieval-instruct2 & Llama-3-Instruct & 8B & - & Zero-shot & N & - \\
    NISTRetrieval-reason0 & Llama-3-Instruct & 8B & - & CoT & N & - \\
    NISTRetrieval-reason1 & Llama-3-Instruct & 8B & - & CoT & N & - \\
    NISTRetrieval-reason2 & Llama-3-Instruct & 8B & - & CoT & N & - \\
    Olz-exp & GPT-4o & - & - & Zero-Shot & S & - \\
    Olz-gpt4o & GPT-4o & - & - & CoT & S & -  \\
    Olz-halfbin & Llama-3-Instruct & 8B & - & CoT & S + N & LR \\
    Olz-somebin & Llama-3-Instruct & 8B  & - & CoT & S + N & LR \\
    Olz-multiprompt & Llama-3-Instruct & 8B & - & CoT & S + N & \checkmark \\
    RMITIR-GPT4o & GPT-4o & - & - & Zero-Shot & N & \checkmark \\
    RMITIR-llama38b & Llama-3-Instruct & 8B & - & Zero-Shot & N & \checkmark \\
    RMITIR-llama70B & Llama-3-Instruct & 70B & - & Zero-Shot & N & \checkmark \\
    TREMA-4prompts & Llama-3-Instruct & 8B & - & Zero-Shot & N & - \\
    TREMA-CoT & Llama-3-Instruct & 8B & - & CoT & N & - \\
    TREMA-all & ChatGPT-3.5/FlanT5-Large & 783M & - & Few-Shot & N & ET \\
    TREMA-direct & ChatGPT-3.5/FlanT5-Large & 783M & - & Few-Shot & N & ET \\
    TREMA-naiveBdecompose & ChatGPT-3.5/FlanT5-Large & 783M & - & Zero-Shot & N & GNB \\
    TREMA-nuggets & ChatGPT-3.5/FlanT5-Large & 783M & - & Zero-Shot & N & ET \\
    TREMA-other & ChatGPT-3.5/FlanT5-Large & 783M & - & Zero-Shot & N & - \\
    TREMA-questions & ChatGPT-3.5/FlanT5-Large & 783M & - & Zero-Shot & N & ET \\
    TREMA-rubric0 & ChatGPT-3.5/FlanT5-Large & 783M & - & Zero-Shot & N & - \\
    TREMA-sumdecompose & Llama-3-Instruct & 8B & - & Zero-Shot & N & - \\
    h2oloo-fewself & GPT-4o & - & - & Few-Shot & N & - \\
    h2oloo-zeroshot1 & Llama-3-Instruct & 8B & \checkmark & Zero-Shot & N & - \\
    h2oloo-zeroshot2 & Llama-3-Instruct & 8B & \checkmark & Zero-Shot & N & - \\
    llmjudge-cot1 & GPT-3.5-turbo & - & - & CoT & N & - \\
    llmjudge-cot2 & GPT-3.5-turbo-16k & - & - & CoT & N & - \\
    llmjudge-cot3 & GPT-4-32k & - & - & CoT & N & - \\
    llmjudge-simple1 & GPT-3.5-turbo & - & - & Zero-Shot & N & - \\
    llmjudge-simple2 & GPT-3.5-turbo-16k & - & - & Zero-Shot & N & - \\
    llmjudge-simple3 & GPT-4-32k & - & - & Zero-shot & N & - \\
    llmjudge-thomas1 & GPT-3.5-turbo & - & - & Zero-Shot & N & - \\
    llmjudge-thomas2 & GPT-3.5-turbo-16k & - & - & Zero-Shot & N & - \\
    llmjudge-thomas3 & GPT-4-32k & - & - & Zero-Shot & N & - \\
    prophet-setting1 & Llama-3-Instruct & 8B & \checkmark & Zero-Shot & S & - \\
    prophet-setting2 & Llama-3-Instruct & 8B & \checkmark & Zero-Shot & S & - \\
    prophet-setting4 & Llama-3-Instruct & 8B & \checkmark & Zero-Shot & S & - \\
    willia-umbrela1 & GPT-4o & - & - & Zero-Shot & N & - \\
    willia-umbrela2 & GPT-4o & - & - & Zero-Shot & S & - \\
    willia-umbrela3 & GPT-4o & - & - & Zero-Shot & S + N & \checkmark \\
    \bottomrule
    \end{tabular}
    }
    \label{tab:infos}
\end{table}

%% file: tables/tbl_results.tex
\begin{table}[]
    \centering
    \caption{Judgment and system ranking correlation of LLMJudge submissions. $\kappa$: Cohen's Kappa, $\alpha$: Krippendorff's alpha, $\tau$: Kendall's Tau, $\rho$: Spearman’s rank correlation. The best results per column are denoted in bold and the second best results are denoted in \textit{italic}.}
    \adjustbox{max width=\columnwidth}{%
    \begin{tabular}{lcccc}
    \toprule
    \textbf{Submission ID} & \textbf{$\kappa$} & \textbf{$\alpha$} & \textbf{$\tau$} & \textbf{$\rho$} \\
    \midrule
    NISTRetrieval-instruct0 & 0.1877 & 0.3819 & 0.9440 & 0.9907 \\
    NISTRetrieval-instruct1 & 0.1874 & 0.3812 & 0.9440 & 0.9907 \\
    NISTRetrieval-instruct2 & 0.1880 & 0.3821 & 0.9440 & 0.9907 \\
    NISTRetrieval-reason0   & 0.1844 & 0.3874 & 0.9052 & 0.9810 \\
    NISTRetrieval-reason1   & 0.1845 & 0.3872 & 0.9009 & 0.9802 \\
    NISTRetrieval-reason2   & 0.1838 & 0.3874 & 0.9052 & 0.9810 \\
    Olz-exp                 & 0.2519 & 0.4701 & 0.9009 & 0.9819 \\
    Olz-gpt4o               & 0.2625 & 0.5020 & 0.8793 & 0.9758 \\
    Olz-halfbin             & 0.2064 & 0.4536 & 0.9085 & 0.9830 \\
    Olz-multiprompt         & 0.2445 & 0.4551 & 0.9267 & 0.9867 \\
    Olz-somebin             & 0.2109 & 0.4471 & 0.9042 & 0.9822 \\
    RMITIR-GPT4o            & 0.2388 & 0.4108 & 0.8966 & 0.9798 \\
    RMITIR-llama38b         & 0.2006 & 0.3873 & 0.8879 & 0.9758 \\
    RMITIR-llama70B         & 0.2654 & 0.4873 & 0.9353 & 0.9883 \\
    TREMA-4prompts          & 0.1829 & 0.2888 & \textit{0.9483} & \textbf{0.9919} \\
    TREMA-CoT               & 0.1961 & 0.3852 & 0.8956 & 0.9799 \\
    TREMA-all               & 0.1471 & 0.3855 & 0.9138 & 0.9863 \\
    TREMA-direct            & 0.1742 & 0.3729 & 0.9009 & 0.9819 \\
    TREMA-naiveBdecompose   & 0.1741 & 0.3579 & 0.9128 & 0.9838 \\
    TREMA-nuggets           & 0.0604 & 0.1691 & 0.8664 & 0.9718 \\
    TREMA-other             & 0.1408 & 0.2712 & 0.8276 & 0.9447 \\
    TREMA-questions         & 0.1137 & 0.3148 & 0.9095 & 0.9839 \\
    TREMA-rubric0           & 0.0779 & 0.1036 & 0.8276 & 0.9544 \\
    TREMA-sumdecompose      & 0.2088 & 0.3926 & 0.9300 & 0.9870 \\
    h2oloo-fewself          & 0.2774 & \textbf{0.4958} & 0.9085 & 0.9822 \\
    h2oloo-zeroshot1        & \textit{0.2817} & 0.4812 & 0.9181 & 0.9827 \\
    h2oloo-zeroshot2        & 0.2589 & 0.3898 & 0.8353 & 0.9604 \\
    llmjudge-cot1           & 0.1284 & 0.3218 & 0.9267 & 0.9871 \\
    llmjudge-cot2           & 0.1560 & 0.3263 & 0.9267 & 0.9875 \\
    llmjudge-cot3           & 0.2271 & 0.4870 & 0.9267 & 0.9851 \\
    llmjudge-simple1        & 0.0754 & 0.2808 & 0.9181 & 0.9863 \\
    llmjudge-simple2        & 0.1327 & 0.3672 & 0.8966 & 0.9790 \\
    llmjudge-simple3        & 0.2110 & 0.4642 & 0.9052 & 0.9810 \\
    llmjudge-thomas1        & 0.1236 & 0.3207 & 0.8664 & 0.9689 \\
    llmjudge-thomas2        & 0.1723 & 0.3853 & 0.8793 & 0.9750 \\
    llmjudge-thomas3        & 0.2293 & 0.4877 & 0.9181 & 0.9867 \\
    prophet-setting1        & 0.1823 & 0.4069 & 0.9042 & 0.9826 \\
    prophet-setting2        & 0.1757 & 0.3144 & \textbf{0.9516} & \textit{0.9914} \\
    prophet-setting4        & 0.1471 & 0.1623 & 0.8568 & 0.9608 \\
    willia-umbrela1         & \textbf{0.2863} & \textit{0.4918} & 0.9009 & 0.9806 \\
    willia-umbrela2         & 0.2688 & 0.4556 & 0.8870 & 0.9769 \\
    willia-umbrela3         & 0.2741 & 0.4535 & 0.8707 & 0.9730 \\
    \bottomrule
    \end{tabular}
    }
    
    \label{tab:main-results}
\end{table}

%% file: sections/05-results.tex
\section{Results}
\label{sec:results}

\input{tables/tbl_kappa}

\subsection{Descriptive Statistics}
For the LLMJudge challenge, we received a total of 42 submissions (i.e., the 42 labelers) from 8 teams. Of these, 9 runs were from the organizers, who contributed a range of baselines; these are included in the statistics. Among the submissions, 5 utilized fine-tuning for relevance judgment, while the rest relied on various prompting techniques. Moreover, 16 submissions are based on proprietary LLMs, whereas 26 used open-source LLMs.

\subsection{Methodological Comparison}
Here we provide an overview analysis of the submissions based on different methodological directions they applied for their LLM-based relevance assessments. Table \ref{tab:main-results} presents the results of submissions at the label and system ranking correlation. Comparing submissions concerning the models they used, we can see that while \texttt{willia-umbrela1} (GPT-4o) achieves the best Cohen's $\kappa$, \texttt{h2oloo-zeroshot1} (Llama3-8B) ranked second by only 1.61\% differences. More interestingly, when we compare the system ranking correlation (i.e., Kendall's $\tau$) of submissions, the best non-fine-tuned method is \texttt{TREMA-4prompts} which uses Llama3-8B. Previous studies \cite{thomas2023large,rahmani2024judgeblender} have shown the importance of the prompting techniques for automatic relevance judgments, and analyzing the LLMJudge submission results confirms the importance of the effect of prompt engineering. For instance, \texttt{TREMA-4prompts} uses the criteria decomposition technique by breaking down the concept of relevance into various criteria and generating the relevance label by asking the model about the specified criteria. Few submissions used fine-tuning, however, their results show that fine-tuning can lead to the highest correlation. For instance, \texttt{h2oloo-fewself} achieved the highest Krippendorff's $\alpha$ (0.4958, the best among all methods) and \texttt{prophet-setting2} is the best-performing submission considering Kendall's $\tau$. This confirms that fine-tuning can significantly enhance agreement with human judgments. Submissions that included both numerical and semantic labels tend to perform consistently across all evaluation metrics compared to those using only numerical labels. For example, submissions from the ``Olz'' team rank among the comparable submissions across all four evaluation metrics. In the following, we provide a more detailed discussion and analysis.

\begin{figure}
    \centering
    \includegraphics[width=0.8\linewidth]{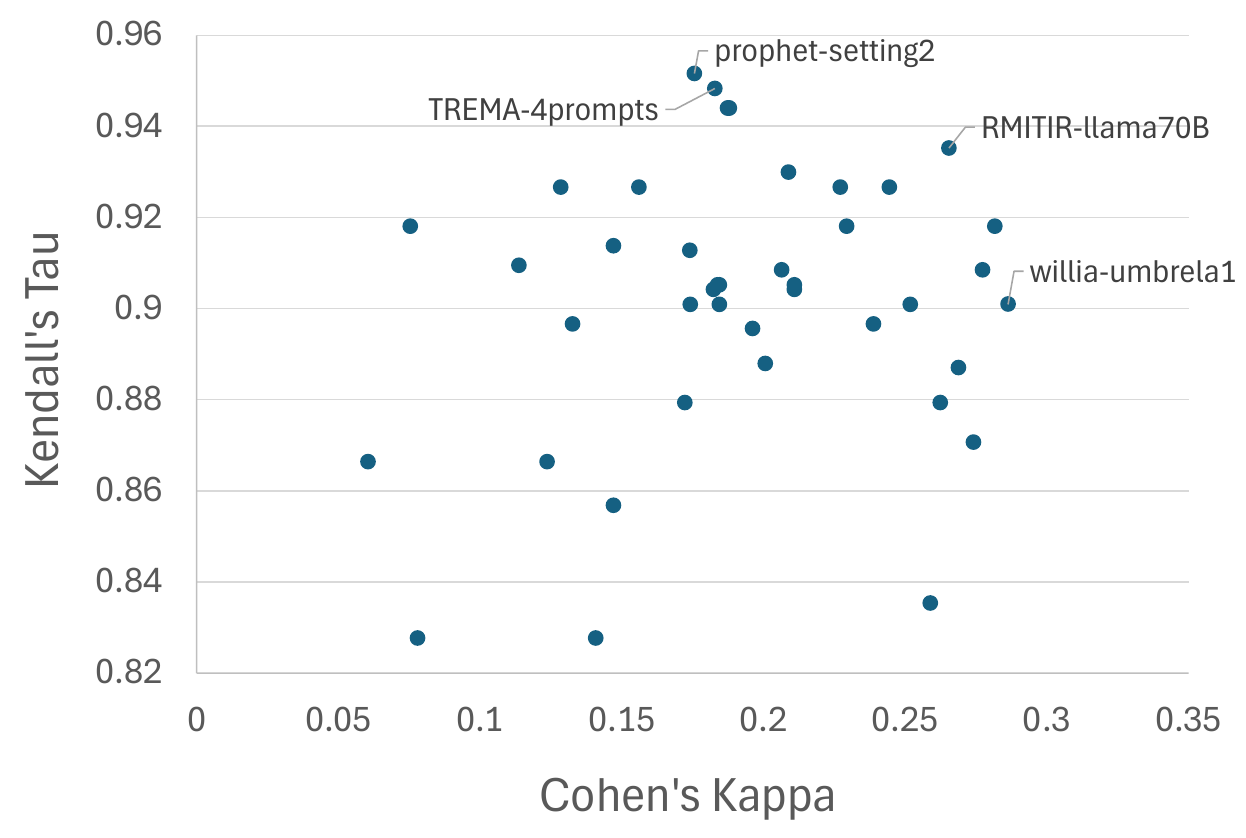}
    \caption{Cohen's $\kappa$ vs.~Kendall's $\tau$}
    \label{fig:kappa-vs-tau}
\end{figure}

\subsection{Overall Results}
The results of the LLMJudge challenge, as presented in Table \ref{tab:main-results}, reveal significant variability in performance across the evaluated metrics, including Cohen's Kappa ($\kappa$), Krippendorff's Alpha ($\alpha$), Kendall's Tau ($\tau$), and Spearman’s Rho ($\rho$). Submissions such as \texttt{h2oloo-fewself}, \texttt{h2oloo-zeroshot1}, and \texttt{willia-umbrela1} emerge as top performers, achieving $\kappa$ values above $0.27$ and $\alpha$ values around $0.49$, indicative of strong agreement and reliability. These methods also demonstrate robust ranking capabilities, with $\tau$ and $\rho$ values exceeding $0.9$. This combination of high agreement and ranking correlation suggests that these approaches are well-suited for relevance judgment tasks. In contrast, submissions like \texttt{TREMA-nuggets} and \texttt{TREMA-rubric0} perform poorly, with $\kappa < 0.1$ and $\alpha < 0.2$, reflecting low agreement and reliability. Despite their low scores on agreement metrics, some of these models maintain moderate ranking correlations, indicating limited but specific utility in ranking-focused scenarios.

A comparison of group trends highlights the strengths and weaknesses of different methodological approaches. For instance, \texttt{NISTRetrieval} submissions consistently achieve high $\tau$ and $\rho$ values ($>0.9$), reflecting strong ranking performance, yet their lower $\kappa$ (\textasciitilde0.18) and $\alpha$ (\textasciitilde0.38) suggest limited alignment with human relevance judgments. In contrast, methods like \texttt{Olz-multiprompt} and \texttt{h2oloo-fewself} demonstrate comparable performance across all metrics. \texttt{Olz-multiprompt} leverages outputs from multiple prompts and \texttt{h2oloo-fewself} incorporates few-shot examples using a proprietary model, GPT4o, these approaches effectively mitigate individual model biases and enhance both reliability and agreement. On the other hand, single-model approaches, such as \texttt{TREMA-direct} and \texttt{llmjudge-simple1}, exhibit limited performance, underscoring the challenges faced by individual models in capturing the complexity of relevance judgment tasks.

\subsection{Label Correlation (Cohen's $\kappa$)}
Table \ref{tab:kappa} presents Cohen's $\kappa$ values for various submissions, providing insights into the agreement between relevance judgments under different granularity levels: 4-point, 0|123, 01|23, and 012|3. Across all groupings, there is noticeable variability in $\kappa$ scores, highlighting differences in consistency among submissions. Submissions like \texttt{h2oloo-fewsel} and \texttt{h2oloo-zeroshot1} demonstrate relatively high agreement across all groupings, particularly excelling in the 4-point and binary (0|123) categories, with $\kappa$ values exceeding 0.27 in most cases. In contrast, submissions such as \texttt{TREMA-nuggets} and \texttt{TREMA-rubric0} exhibit significantly lower agreement, with $\kappa$ values as low as 0.0604 and 0.0779 on the 4-point scale, reflecting limited reliability in their judgments.

In particular, coarse-grained groupings like 0|123 tend to produce higher $\kappa$ values than finer groupings like the 4-point scale, suggesting that systems or annotators achieve better consistency when relevance levels are aggregated. However, the 012|3 grouping, which isolates the highest relevance level, introduces greater variability, with some systems such as \texttt{willia-umbrela1} performing well, while others struggle to maintain consistency. These findings emphasize the importance of evaluation granularity in understanding system reliability and identifying approaches with stable performance across diverse grouping strategies.

\subsection{Cohen's $\kappa$ vs.~Kendall's $\tau$}
Figure \ref{fig:kappa-vs-tau} shows the performance of submitted runs on the \texttt{LLMJudge} test set. The x-axis represents Cohen's $\kappa$, and the y-axis shows the submission agreement on system order. Submissions exhibit low variability in Kendall's $\tau$ but greater variability in Cohen's $\kappa$. Most submissions cluster within a narrow range of $\tau$ values, indicating consistent system rankings but more variation in inter-rater reliability, as measured by Cohen's $\kappa$. This suggests that while submissions generally agree on rankings, their exact labels are less consistent, leading to the observed variability in $\kappa$.

\subsection{Average Label Comparison}

\begin{figure}
    \centering
    \subfloat[Avg.~Label vs.~Kendall $\tau$\label{fig:avg-labels-1}]
    {
        {
            \includegraphics[scale=0.2]{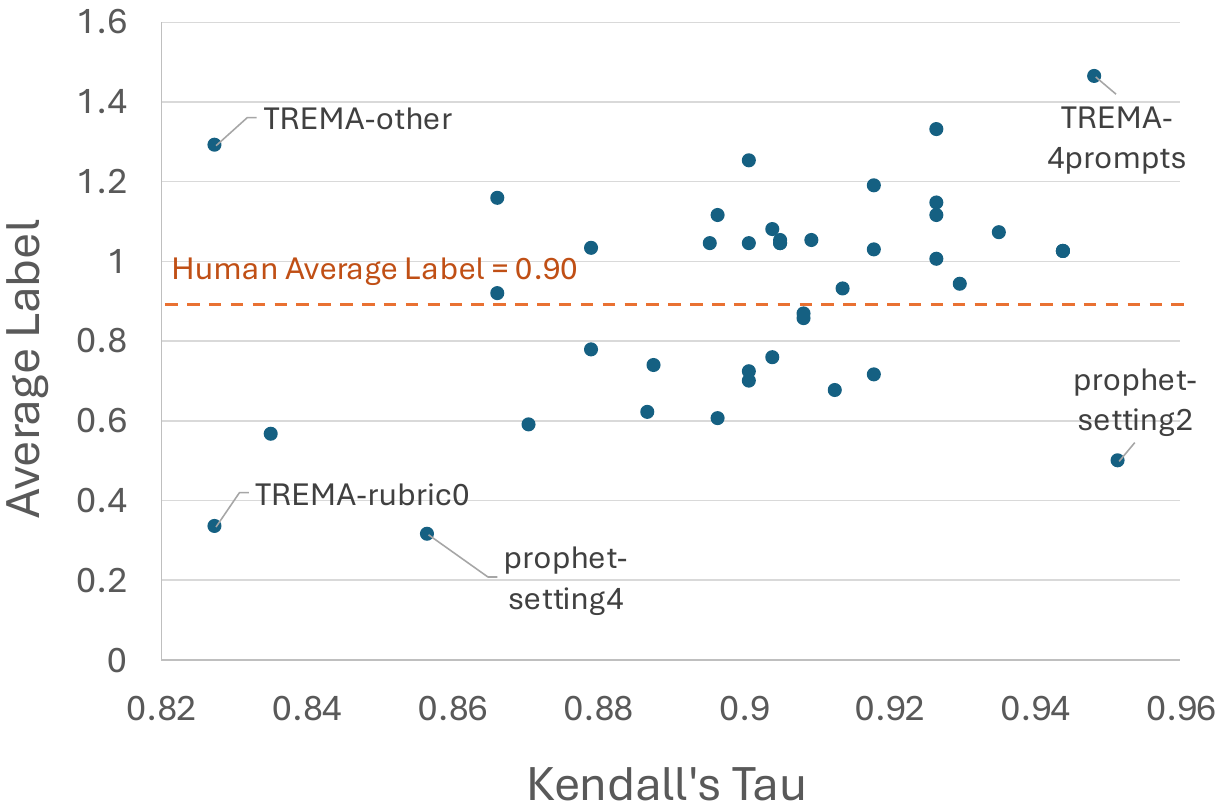}
        }
    }%
    \subfloat[Avg.~Label vs.~Krippendorff's $\alpha$\label{fig:avg-labels-2}]
    {
        {
            \includegraphics[scale=0.2]{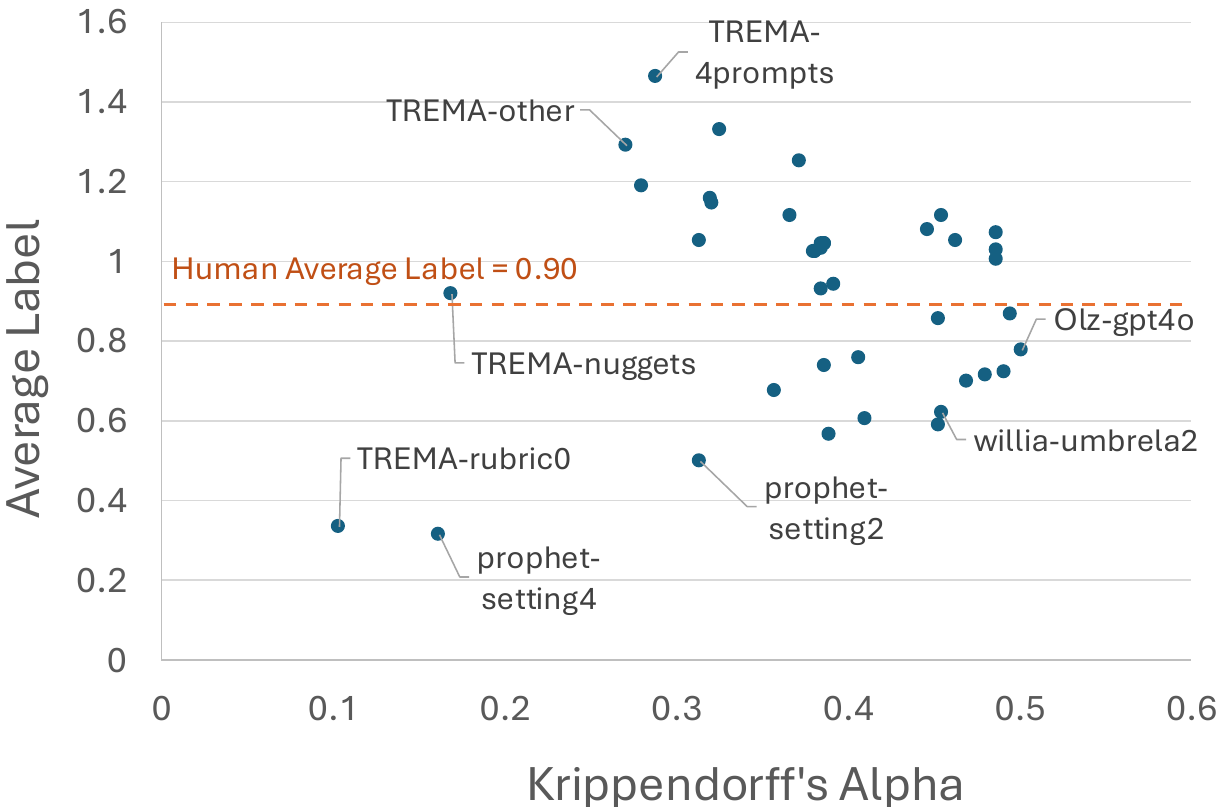}
        }
    }%
    \caption{Average Labels}%
    \label{fig:avg-labels}%
\end{figure}

Figure \ref{fig:avg-labels} illustrates the relationship between ranking correlation metrics and the average label assigned by different evaluation methods for NDCG@10. The \textcolor{orange}{orange} dashed line represents the human average label (0.90), serving as a baseline for comparison. In Figure \ref{fig:avg-labels-1}, most methods exhibit strong agreement in ranking order, with values generally above 0.85. However, the assigned average labels vary significantly, with some methods, such as \texttt{TREMA-other} and \texttt{TREMA-4prompts}, assigning scores notably above the human baseline, while others, like \texttt{TREMA-rubric0} and \texttt{prophet-setting4}, produce lower scores. These variations indicate that while many approaches maintain ranking consistency, their absolute scoring tendencies differ, potentially introducing biases in evaluation.

Figure \ref{fig:avg-labels-2} provides further insight into inter-method agreement, capturing not just ranking order but also overall consistency in score distributions. Here, we observe a wider range of correlation values, with some methods achieving moderate agreement (e.g., \texttt{Olz-gpt4o} and \texttt{willia-umbrella2}) while others, such as \texttt{TREMA-rubric0} and \texttt{prophet-setting4}, show very low agreement and lower assigned scores. Notably, \texttt{TREMA-4prompts} and \texttt{TREMA-\\other} again stand out with higher average labels, but their variability suggests differences in how they align with human judgments. These findings emphasize the importance of calibration when aggregating synthetic judgments, as different methods may systematically overestimate or underestimate relevance scores despite high-ranking agreement.

\begin{figure}
    \centering
    \includegraphics[width=0.8\linewidth]{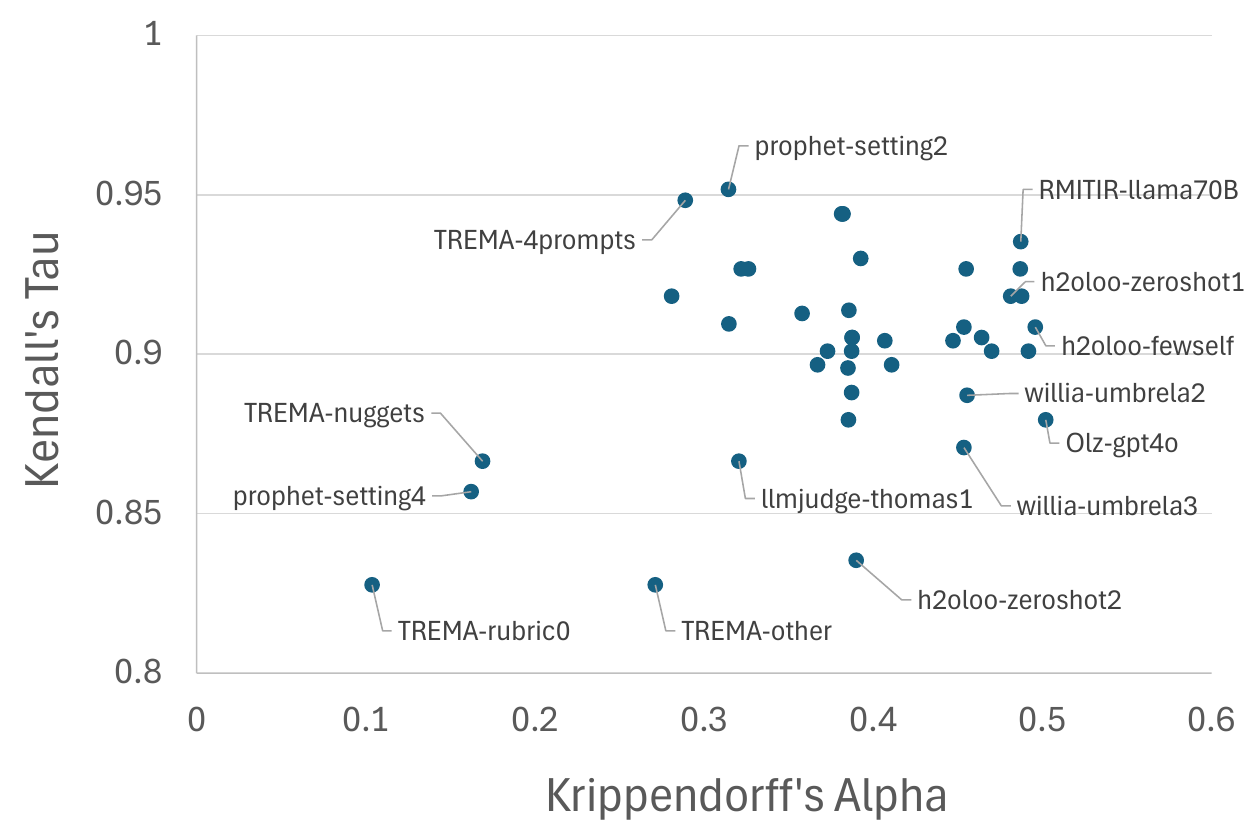}
    \caption{Krippendorff's $\alpha$ vs.~Kendall's $\tau$}
    \label{fig:alpha-vs-tau}
\end{figure}

\subsection{Krippendorff's \texorpdfstring{$\alpha$}{alpha} vs.~Kendall's \texorpdfstring{$\tau$}{tau}}
Figure \ref{fig:alpha-vs-tau} shows the relation between Krippendorff's $\alpha$ and Kendall's $\tau$, highlighting the agreement of different models with human judgments. Higher Krippendorff's $\alpha$ generally corresponds to better Kendall’s $\tau$, but with notable variance. Models like \texttt{prophet-setting2} and \texttt{TREMA-4prompts} achieve strong ranking consistency, while \texttt{TREMA-rubric0} and \texttt{TREMA-other} show weaker agreement. Proprietary models such as \texttt{h2oloo-fewself} and \texttt{Olz-gpt4o} perform well in both metrics, whereas some open-source models are more dispersed.

\subsection{Binarized Cohen's $\kappa$ vs.~Kendall's $\tau$}
\begin{figure}
    \centering
    \includegraphics[width=0.8\linewidth]{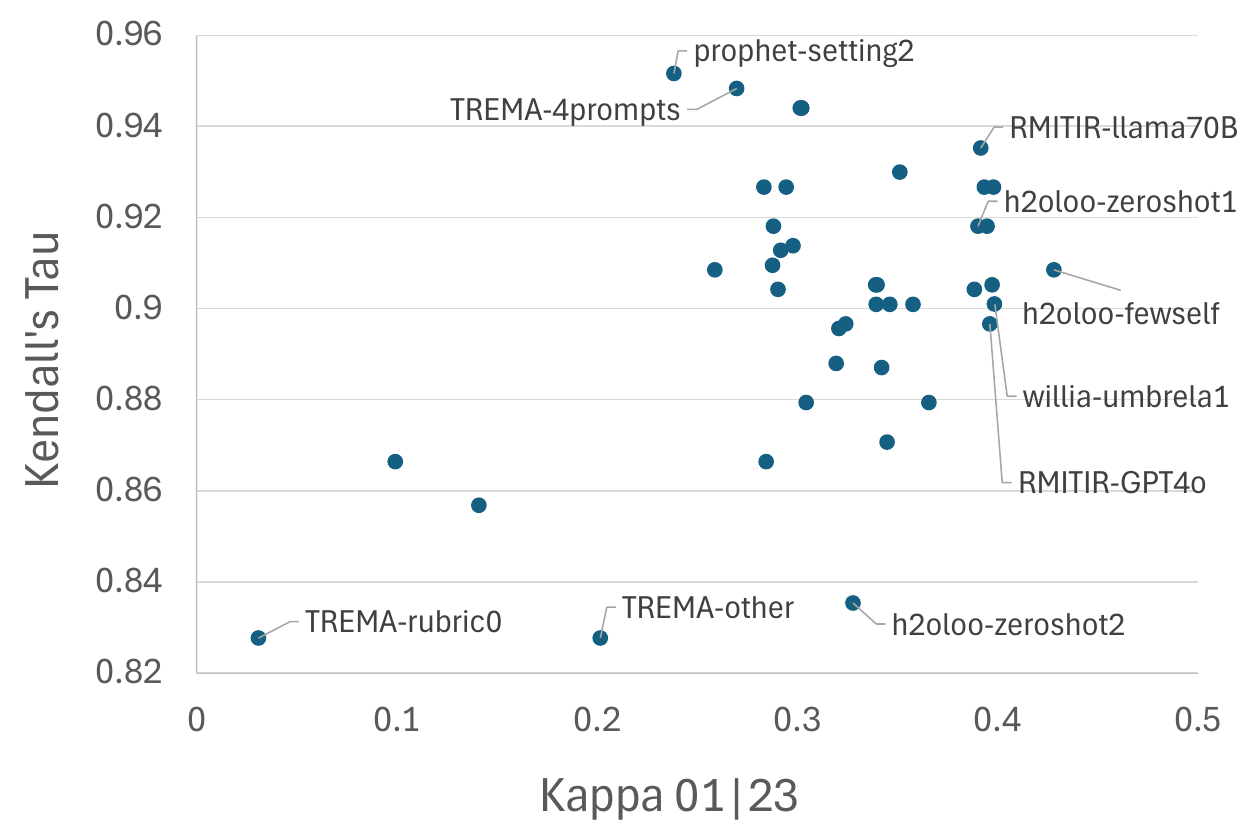}
    \caption{Binarized Cohen's $\kappa$ vs.~Kendall's $\tau$}
    \label{fig:binkappa-vs-tau}
\end{figure}

Figure \ref{fig:binkappa-vs-tau} compares binary agreement (Cohen's $\kappa$ 01|23) with Kendall's $\tau$ across models. Higher kappa values tend to align with stronger ranking consistency, as seen with \texttt{prophet-setting2}, \texttt{TREMA-\\4prompts}, and \texttt{RMITIR-llama70B}. However, some models, like \texttt{TREMA-\\rubric0} and \texttt{TREMA-other}, show weak agreement despite moderate ranking correlations. Proprietary models, including \texttt{RMITIR-GPT4o} and \texttt{h2oloo-fewself}, perform competitively, suggesting that both fine-tuning and prompting strategies impact these relationships.

\subsection{LLMJudge Resource Use Cases}
The LLMJudge benchmark can be considered as a resource for evaluating the reliability of LLM-generated relevance judgments across different settings. For instance, JudgeBlender \cite{rahmani2024judgeblender} which is a framework that aggregates evaluations from multiple smaller models to enhance the robustness of relevance assessments recently leveraged LLMJudge both as a baseline for comparison and as a tool for analyzing the variability and bias of ensemble-based judgment aggregation methods.

%% file: tables/tbl_kappa.tex
\begin{table}[]
    \centering
    \caption{Cohen's $\kappa$ correlation in 4-point scale agreement and difference binarize the judgment scale. Judgment levels to the left of the pipe are considered irrelevant, while those to the right are considered relevant.}
    \adjustbox{max width=\columnwidth}{%
    \begin{tabular}{lcccc}
    \toprule
    \textbf{Submission ID} & \textbf{4-point} & \textbf{0|123} & \textbf{01|23} & \textbf{012|3} \\
    \midrule
    NISTRetrieval-instruct0 & 0.1877 & 0.3116 & 0.3021 & 0.0000 \\
    NISTRetrieval-instruct1 & 0.1874 & 0.3106 & 0.3021 & 0.0000 \\
    NISTRetrieval-instruct2 & 0.1880 & 0.3126 & 0.3013 & 0.0000 \\
    NISTRetrieval-reason0   & 0.1844 & 0.2911 & 0.3390 & 0.0097 \\
    NISTRetrieval-reason1   & 0.1845 & 0.2906 & 0.3394 & 0.0097 \\
    NISTRetrieval-reason2   & 0.1838 & 0.2902 & 0.3397 & 0.0097 \\
    Olz-exp                 & 0.2519 & 0.3997 & 0.3577 & 0.2936 \\
    Olz-gpt4o               & 0.2625 & \textbf{0.4228} & 0.3657 & 0.3066 \\
    Olz-halfbin             & 0.2064 & 0.4008 & 0.2587 & 0.2449 \\
    Olz-multiprompt         & 0.2445 & 0.3764 & 0.3934 & 0.2150 \\
    Olz-somebin             & 0.2109 & 0.3854 & 0.3883 & 0.1137 \\
    RMITIR-GPT4o            & 0.2388 & 0.3499 & 0.3961 & 0.2580 \\
    RMITIR-llama38b         & 0.2006 & 0.3280 & 0.3194 & 0.1344 \\
    RMITIR-llama70B         & 0.2654 & \textit{0.4166} & 0.3916 & 0.2843 \\
    TREMA-4prompts          & 0.1829 & 0.3022 & 0.2697 & 0.1664 \\
    TREMA-CoT               & 0.1961 & 0.3181 & 0.3208 & 0.1836 \\
    TREMA-all               & 0.1471 & 0.3244 & 0.2978 & 0.0717 \\
    TREMA-direct            & 0.1742 & 0.3205 & 0.3462 & 0.1763 \\
    TREMA-naiveBdecompose   & 0.1741 & 0.3085 & 0.2916 & 0.0153 \\
    TREMA-nuggets           & 0.0604 & 0.1505 & 0.0992 & -0.0077 \\
    TREMA-other             & 0.1408 & 0.2740 & 0.2015 & 0.1411 \\
    TREMA-questions         & 0.1137 & 0.2636 & 0.2876 & 0.0441 \\
    TREMA-rubric0           & 0.0779 & 0.1714 & 0.0308 & 0.0369 \\
    TREMA-sumdecompose      & 0.2088 & 0.3228 & 0.3512 & 0.2047 \\
    h2oloo-fewself          & 0.2774 & 0.4172 & 0.4280 & 0.3048 \\
    h2oloo-zeroshot1        & 0.2817 & 0.4094 & 0.3901 & 0.3084 \\
    h2oloo-zeroshot2        & 0.2589 & 0.3691 & 0.3278 & 0.2789 \\
    llmjudge-cot1           & 0.1284 & 0.2219 & 0.2833 & 0.1287 \\
    llmjudge-cot2           & 0.1560 & 0.2507 & 0.2944 & 0.2048 \\
    llmjudge-cot3           & 0.2271 & 0.3856 & 0.3978 & 0.2335 \\
    llmjudge-simple1        & 0.0754 & 0.1278 & 0.2880 & 0.1582 \\
    llmjudge-simple2        & 0.1327 & 0.2349 & 0.3241 & 0.2004 \\
    llmjudge-simple3        & 0.2110 & 0.3590 & 0.3972 & 0.2157 \\
    llmjudge-thomas1        & 0.1236 & 0.2087 & 0.2843 & 0.1802 \\
    llmjudge-thomas2        & 0.1723 & 0.2944 & 0.3043 & 0.2267 \\
    llmjudge-thomas3        & 0.2293 & 0.3910 & 0.3947 & 0.2438 \\
    prophet-setting1        & 0.1823 & 0.3502 & 0.2903 & 0.1677 \\
    prophet-setting2        & 0.1757 & 0.3102 & 0.2382 & 0.0284 \\
    prophet-setting4        & 0.1471 & 0.2375 & 0.1409 & 0.0371 \\
    willia-umbrela1         & 0.2863 & 0.4161 & \textbf{0.3985} & 0.3145 \\
    willia-umbrela2         & 0.2688 & 0.4109 & 0.3421 & 0.3194 \\
    willia-umbrela3         & 0.2741 & 0.4114 & 0.3447 & \textbf{0.3215} \\
    \bottomrule
    \end{tabular}
    }
    
    \label{tab:kappa}
\end{table}

%% file: sections/06-conclusion.tex
\section{Conclusion}
\label{sec:conclusion}
We introduce the LLMJudge resource, which builds upon the foundations established by the LLMJudge challenge \cite{rahmani2024llmjudge} at the LLM4Eval workshop \cite{rahmani2024llm4eval,rahmani2024report} co-located with SIGIR 2024. This resource provides a benchmark for evaluating the effectiveness of LLMs in an automatic relevance judgment task, helping comparisons across different models and prompting strategies.
In this paper, we detail the 42 sets of relevance judgments for the TREC 2023 Deep Learning track submitted to the LLMJudge challenge by 8 different international teams. 
We release this collection to serve multiple purposes. First, they can be used as a comparison for future LLM-based relevance assessments, allowing research teams that did not participate in the challenge to compare their approaches as well. Secondly, this resource can serve as a tool to determine or help empirically study the presence of systematic biases in LLM-generated relevance judgments, impacting a large body of research in the community.
Among the submitted sets of relevance judgments, 5 employ fine-tuning, and their results show that fine-tuning can lead to the highest correlation. Furthermore, 16 submissions are based on proprietary LLMs and 26 on open-source LLMs. Our analyses show that the latter tend to be more stable, while the former are affected by higher variability.
Methodologically, we provide in this paper a set of strategies to compare multiple automatically-generated relevance assessments that can serve future practitioners in determining the effectiveness of new LLMs as assessors.
In future work, we plan to investigate how LLM can be used to generate relevance judgment in a nugget-based evaluation scenario and extend the analysis to fully automatic collections, that include automatically generated queries and documents.

%% file: tables/tbl_labels.tex
\begin{table}[ht]
    \centering
    \caption{The number of labels assigned by human judges and LLMJudge challenge submissions to each judgment level. Bold indicates the closest prediction to the number of labels assigned by humans.}
    \adjustbox{max width=\columnwidth}{%
    \begin{tabular}{lcccc}
    \toprule
    \textbf{Submission ID} & \textbf{0} & \textbf{1} & \textbf{2} & \textbf{3} \\
    \midrule
    human & 2005 & 1233 & 808 & 377 \\
    \midrule
    NISTRetrieval-instruct0 & 1115 & 2092 & 1216 & 0 \\
    NISTRetrieval-instruct1 & 1115 & 2092 & 1216 & 0 \\
    NISTRetrieval-instruct2 & 1117 & 2088 & 1218 & 0 \\
    NISTRetrieval-reason0 & 1159 & 1922 & 1340 & 2 \\
    NISTRetrieval-reason1 & 1158 & 1924 & 1339 & 2 \\
    NISTRetrieval-reason2 & 1159 & 1921 & 1341 & 2 \\
    Olz-exp & 2435 & 1210 & 456 & 322 \\
    Olz-gpt4o & 2258 & 1274 & 504 & \textbf{387} \\
    Olz-halfbin & 2100 & 1458 & 277 & 588 \\
    Olz-multiprompt & 1713 & 976 & 1244 & 490 \\
    Olz-somebin & 2102 & 595 & 1004 & 722 \\
    RMITIR-GPT4o & 3056 & 349 & 730 & 288 \\
    RMITIR-llama38b & 2576 & 614 & 1058 & 175 \\
    RMITIR-llama70B & 2154 & 243 & 1581 & 443 \\
    TREMA-4prompts & 1027 & 751 & 2213 & 432 \\
    TREMA-CoT & 1835 & 1122 & 906 & 560 \\
    TREMA-all & 2399 & 616 & 734 & 674 \\
    TREMA-direct & 2404 & 87 & 342 & 1590 \\
    TREMA-naiveBdecompose & 2627 & 645 & 1117 & 34 \\
    TREMA-nuggets & 2127 & 893 & 1044 & 359 \\
    TREMA-other & 1305 & 809 & 2021 & 288 \\
    TREMA-questions & 2450 & 253 & \textbf{767} & 953 \\
    TREMA-rubric0 & 3122 & 1211 & 0 & 90 \\
    TREMA-sumdecompose & 2518 & 254 & 1049 & 602 \\
    h2oloo-fewself & 2470 & 732 & 557 & 664 \\
    h2oloo-zeroshot1 & 2353 & 1225 & 597 & 248 \\
    h2oloo-zeroshot2 & 2920 & 771 & 476 & 255 \\
    llmjudge-cot1 & 991 & 1921 & 1383 & 128 \\
    llmjudge-cot2 & 1111 & 955 & 2149 & 208 \\
    llmjudge-cot3 & 1902 & 1321 & 486 & 714 \\
    llmjudge-simple1 & 777 & 2228 & 1217 & 200 \\
    llmjudge-simple2 & 1720 & 905 & 1375 & 423 \\
    llmjudge-simple3 & 1834 & 1318 & 485 & 786 \\
    llmjudge-thomas1 & 1049 & 1803 & 1402 & 169 \\
    llmjudge-thomas2 & 1886 & 733 & 1585 & 219 \\
    llmjudge-thomas3 & \textbf{1934} & 1184 & 559 & 746 \\
    prophet-setting1 & 2506 & 885 & 629 & 403 \\
    prophet-setting2 & 2903 & 852 & 651 & 17 \\
    prophet-setting4 & 3359 & 763 & 281 & 20 \\
    willia-umbrela1 & 2335 & \textbf{1231} & 608 & 249 \\
    willia-umbrela2 & 2705 & 1029 & 350 & 339 \\
    willia-umbrela3 & 2710 & 1041 & 446 & 226 \\
    \bottomrule
    \end{tabular}
    }
    \label{tab:labels}
\end{table}

%% file: tables/tbl_alpha.tex
\begin{table}[ht]
    \centering
    \caption{Krippendorff's $\alpha$ correlation in 4-point scale agreement and difference binarize the judgment scale. Judgment levels to the left of the pipe are considered irrelevant, while those to the right are considered relevant.}
    \adjustbox{max width=\columnwidth}{%
    \begin{tabular}{lcccc}
    \toprule
    \textbf{Submission ID} & \textbf{4-point} & \textbf{0|123} & \textbf{01|23} & \textbf{012|3} \\
    \midrule
    NISTRetrieval-instruct0 & 0.3819 & 0.2811 & 0.3021 & -0.0444 \\
    NISTRetrieval-instruct1 & 0.3812 & 0.2801 & 0.3021 & -0.0444 \\
    NISTRetrieval-instruct2 & 0.3821 & 0.2823 & 0.3013 & -0.0444 \\
    NISTRetrieval-reason0 & 0.3874 & 0.263 & 0.3381 & -0.0336 \\
    NISTRetrieval-reason1 & 0.3872 & 0.2624 & 0.3385 & -0.0336 \\
    NISTRetrieval-reason2 & 0.3874 & 0.262 & 0.3388 & -0.0336 \\
    Olz-exp & 0.4701 & 0.3941 & 0.3499 & 0.2933 \\
    Olz-gpt4o & 0.502 & 0.421 & 0.3619 & 0.3067 \\
    Olz-halfbin & 0.4536 & 0.4005 & 0.2534 & 0.2405 \\
    Olz-multiprompt & 0.4551 & 0.3737 & 0.3829 & 0.2137 \\
    Olz-somebin & 0.4471 & 0.3851 & 0.378 & 0.1014 \\
    RMITIR-GPT4o & 0.4108 & 0.3125 & 0.395 & 0.257 \\
    RMITIR-llama38b & 0.3873 & 0.3169 & 0.3194 & 0.1268 \\
    RMITIR-llama70B & 0.4873 & 0.416 & 0.3679 & 0.2839 \\
    TREMA-4prompts & 0.2888 & 0.2644 & 0.1888 & 0.1661 \\
    TREMA-CoT & 0.3852 & 0.3172 & 0.3176 & 0.18 \\
    TREMA-all & 0.3855 & 0.3191 & 0.2957 & 0.0618 \\
    TREMA-direct & 0.3729 & 0.315 & 0.3259 & 0.0868 \\
    TREMA-naiveBdecompose & 0.3579 & 0.2949 & 0.2916 & -0.018 \\
    TREMA-nuggets & 0.1691 & 0.1499 & 0.0967 & -0.0076 \\
    TREMA-other & 0.2712 & 0.2547 & 0.1477 & 0.1399 \\
    TREMA-questions & 0.3148 & 0.2562 & 0.2758 & 0.0125 \\
    TREMA-rubric0 & 0.1036 & 0.1172 & -0.0895 & 0.0167 \\
    TREMA-sumdecompose & 0.3926 & 0.3138 & 0.343 & 0.1995 \\
    h2oloo-fewself & 0.4958 & 0.4108 & 0.428 & 0.2978 \\
    h2oloo-zeroshot1 & 0.4812 & 0.4058 & 0.385 & 0.3063 \\
    h2oloo-zeroshot2 & 0.3898 & 0.3418 & 0.3175 & 0.2769 \\
    llmjudge-cot1 & 0.3218 & 0.1764 & 0.2788 & 0.116 \\
    llmjudge-cot2 & 0.3263 & 0.2173 & 0.2429 & 0.2002 \\
    llmjudge-cot3 & 0.487 & 0.3853 & 0.3979 & 0.2233 \\
    llmjudge-simple1 & 0.2808 & 0.05 & 0.2857 & 0.1528 \\
    llmjudge-simple2 & 0.3672 & 0.2317 & 0.3097 & 0.2003 \\
    llmjudge-simple3 & 0.4642 & 0.3581 & 0.397 & 0.2012 \\
    llmjudge-thomas1 & 0.3207 & 0.1679 & 0.278 & 0.1725 \\
    llmjudge-thomas2 & 0.3853 & 0.294 & 0.2891 & 0.2229 \\
    llmjudge-thomas3 & 0.4877 & 0.3909 & 0.3942 & 0.2321 \\
    prophet-setting1 & 0.4069 & 0.3419 & 0.2892 & 0.1677 \\
    prophet-setting2 & 0.3144 & 0.2815 & 0.2225 & -0.0093 \\
    prophet-setting4 & 0.1623 & 0.1627 & 0.0797 & 0.0006 \\
    willia-umbrela1 & 0.4918 & 0.4129 & 0.3939 & 0.3124 \\
    willia-umbrela2 & 0.4556 & 0.3961 & 0.3298 & 0.3193 \\
    willia-umbrela3 & 0.4535 & 0.3965 & 0.3314 & 0.3185 \\
    \bottomrule
    \end{tabular}
    }
    \label{tab:alpha}
\end{table}

%% file: main.bbl

\begin{thebibliography}{38}


\ifx \showCODEN    \undefined \def \showCODEN     #1{\unskip}     \fi
\ifx \showDOI      \undefined \def \showDOI       #1{#1}\fi
\ifx \showISBNx    \undefined \def \showISBNx     #1{\unskip}     \fi
\ifx \showISBNxiii \undefined \def \showISBNxiii  #1{\unskip}     \fi
\ifx \showISSN     \undefined \def \showISSN      #1{\unskip}     \fi
\ifx \showLCCN     \undefined \def \showLCCN      #1{\unskip}     \fi
\ifx \shownote     \undefined \def \shownote      #1{#1}          \fi
\ifx \showarticletitle \undefined \def \showarticletitle #1{#1}   \fi
\ifx \showURL      \undefined \def \showURL       {\relax}        \fi
\providecommand\bibfield[2]{#2}
\providecommand\bibinfo[2]{#2}
\providecommand\natexlab[1]{#1}
\providecommand\showeprint[2][]{arXiv:#2}

\bibitem[Alaofi et~al\mbox{.}(2024)]%
        {DBLP:conf/sigir-ap/Alaofi0SS24}
\bibfield{author}{\bibinfo{person}{Marwah Alaofi}, \bibinfo{person}{Paul Thomas}, \bibinfo{person}{Falk Scholer}, {and} \bibinfo{person}{Mark Sanderson}.} \bibinfo{year}{2024}\natexlab{}.
\newblock \showarticletitle{LLMs can be Fooled into Labelling a Document as Relevant: best caf{\'{e}} near me; this paper is perfectly relevant}. In \bibinfo{booktitle}{\emph{Proceedings of the 2024 Annual International {ACM} {SIGIR} Conference on Research and Development in Information Retrieval in the Asia Pacific Region, {SIGIR-AP} 2024, Tokyo, Japan, December 9-12, 2024}}. \bibinfo{publisher}{{ACM}}, \bibinfo{pages}{32--41}.
\newblock
\urldef\tempurl%
\url{https://doi.org/10.1145/3673791.3698431}
\showDOI{\tempurl}


\bibitem[Aonghusa and Leith(2016)]%
        {DBLP:journals/tissec/AonghusaL16}
\bibfield{author}{\bibinfo{person}{Pol~Mac Aonghusa} {and} \bibinfo{person}{Douglas~J. Leith}.} \bibinfo{year}{2016}\natexlab{}.
\newblock \showarticletitle{Don't Let Google Know I'm Lonely}.
\newblock \bibinfo{journal}{\emph{{ACM} Trans. Priv. Secur.}} \bibinfo{volume}{19}, \bibinfo{number}{1} (\bibinfo{year}{2016}), \bibinfo{pages}{3:1--3:25}.
\newblock
\urldef\tempurl%
\url{https://doi.org/10.1145/2937754}
\showDOI{\tempurl}


\bibitem[Bender et~al\mbox{.}(2021)]%
        {DBLP:conf/fat/BenderGMS21}
\bibfield{author}{\bibinfo{person}{Emily~M. Bender}, \bibinfo{person}{Timnit Gebru}, \bibinfo{person}{Angelina McMillan{-}Major}, {and} \bibinfo{person}{Shmargaret Shmitchell}.} \bibinfo{year}{2021}\natexlab{}.
\newblock \showarticletitle{On the Dangers of Stochastic Parrots: Can Language Models Be Too Big?}. In \bibinfo{booktitle}{\emph{FAccT '21: 2021 {ACM} Conference on Fairness, Accountability, and Transparency, Virtual Event / Toronto, Canada, March 3-10, 2021}}. \bibinfo{publisher}{{ACM}}, \bibinfo{pages}{610--623}.
\newblock
\urldef\tempurl%
\url{https://doi.org/10.1145/3442188.3445922}
\showDOI{\tempurl}


\bibitem[Blanco et~al\mbox{.}(2011)]%
        {DBLP:conf/sigir/BlancoHHMPTT11}
\bibfield{author}{\bibinfo{person}{Roi Blanco}, \bibinfo{person}{Harry Halpin}, \bibinfo{person}{Daniel~M. Herzig}, \bibinfo{person}{Peter Mika}, \bibinfo{person}{Jeffrey Pound}, \bibinfo{person}{Henry~S. Thompson}, {and} \bibinfo{person}{Duc~Thanh Tran}.} \bibinfo{year}{2011}\natexlab{}.
\newblock \showarticletitle{Repeatable and reliable search system evaluation using crowdsourcing}. In \bibinfo{booktitle}{\emph{Proceeding of the 34th International {ACM} {SIGIR} Conference on Research and Development in Information Retrieval, {SIGIR} 2011, Beijing, China, July 25-29, 2011}}. \bibinfo{publisher}{{ACM}}, \bibinfo{pages}{923--932}.
\newblock
\urldef\tempurl%
\url{https://doi.org/10.1145/2009916.2010039}
\showDOI{\tempurl}


\bibitem[Clarke and Dietz(2024)]%
        {DBLP:journals/corr/abs-2412-17156}
\bibfield{author}{\bibinfo{person}{Charles L.~A. Clarke} {and} \bibinfo{person}{Laura Dietz}.} \bibinfo{year}{2024}\natexlab{}.
\newblock \showarticletitle{LLM-based relevance assessment still can't replace human relevance assessment}.
\newblock \bibinfo{journal}{\emph{CoRR}}  \bibinfo{volume}{abs/2412.17156} (\bibinfo{year}{2024}).
\newblock
\urldef\tempurl%
\url{https://doi.org/10.48550/ARXIV.2412.17156}
\showDOI{\tempurl}
\showeprint[arXiv]{2412.17156}


\bibitem[Cleverdon(1967)]%
        {10.5555/275537.275544}
\bibfield{author}{\bibinfo{person}{Cyril Cleverdon}.} \bibinfo{year}{1967}\natexlab{}.
\newblock \bibinfo{booktitle}{\emph{The Cranfield tests on index language devices}}. Vol.~\bibinfo{volume}{19}.
\newblock \bibinfo{publisher}{Emerald}, \bibinfo{address}{San Francisco, CA, USA}, \bibinfo{pages}{173--194}.
\newblock
Issue 6.
\showISBNx{0001-253X}
\urldef\tempurl%
\url{https://doi.org/doi:10.1108/eb050097}
\showDOI{\tempurl}


\bibitem[Cleverdon(1960)]%
        {Cleverdon1960TheAC}
\bibfield{author}{\bibinfo{person}{Cyril~W. Cleverdon}.} \bibinfo{year}{1960}\natexlab{}.
\newblock \showarticletitle{The Aslib Cranfield Research Project on the Comparative Efficiency of Indexing Systems}.
\newblock
\urldef\tempurl%
\url{https://api.semanticscholar.org/CorpusID:60470177}
\showURL{%
\tempurl}


\bibitem[Craswell et~al\mbox{.}(2024)]%
        {craswell2024overview}
\bibfield{author}{\bibinfo{person}{Nick Craswell}, \bibinfo{person}{Bhaskar Mitra}, \bibinfo{person}{Emine Yilmaz}, \bibinfo{person}{Hossein~A. Rahmani}, \bibinfo{person}{Daniel Campos}, \bibinfo{person}{Jimmy Lin}, \bibinfo{person}{Ellen~M. Voorhees}, {and} \bibinfo{person}{Ian Soboroff}.} \bibinfo{year}{2024}\natexlab{}.
\newblock \showarticletitle{Overview of the TREC 2023 Deep Learning Track}. In \bibinfo{booktitle}{\emph{Text REtrieval Conference (TREC)}}. NIST, \bibinfo{publisher}{TREC}.
\newblock


\bibitem[Croft et~al\mbox{.}(2009)]%
        {croft2009search}
\bibfield{author}{\bibinfo{person}{W.~Bruce Croft}, \bibinfo{person}{Donald Metzler}, {and} \bibinfo{person}{Trevor Strohman}.} \bibinfo{year}{2009}\natexlab{}.
\newblock \bibinfo{booktitle}{\emph{Search Engines: Information Retrieval in Practice}}.
\newblock \bibinfo{publisher}{Addison Wesley}.
\newblock
\showISBNx{978-0136072249}
\urldef\tempurl%
\url{https://www.amazon.com/Search-Engines-Information-Retrieval-Practice/dp/0136072240}
\showURL{%
\tempurl}


\bibitem[Dietz(2024a)]%
        {DBLP:conf/sigir/Dietz24}
\bibfield{author}{\bibinfo{person}{Laura Dietz}.} \bibinfo{year}{2024}\natexlab{a}.
\newblock \showarticletitle{A Workbench for Autograding Retrieve/Generate Systems}. In \bibinfo{booktitle}{\emph{Proceedings of the 47th International {ACM} {SIGIR} Conference on Research and Development in Information Retrieval, {SIGIR} 2024, Washington DC, USA, July 14-18, 2024}}. \bibinfo{publisher}{{ACM}}, \bibinfo{pages}{1963--1972}.
\newblock
\urldef\tempurl%
\url{https://doi.org/10.1145/3626772.3657871}
\showDOI{\tempurl}


\bibitem[Dietz(2024b)]%
        {dietz2024workbench}
\bibfield{author}{\bibinfo{person}{Laura Dietz}.} \bibinfo{year}{2024}\natexlab{b}.
\newblock \showarticletitle{A workbench for autograding retrieve/generate systems}. In \bibinfo{booktitle}{\emph{Proceedings of the 47th International ACM SIGIR Conference on Research and Development in Information Retrieval}}. \bibinfo{pages}{1963--1972}.
\newblock


\bibitem[Faggioli et~al\mbox{.}(2023)]%
        {faggioli2023perspectives}
\bibfield{author}{\bibinfo{person}{Guglielmo Faggioli}, \bibinfo{person}{Laura Dietz}, \bibinfo{person}{Charles~LA Clarke}, \bibinfo{person}{Gianluca Demartini}, \bibinfo{person}{Matthias Hagen}, \bibinfo{person}{Claudia Hauff}, \bibinfo{person}{Noriko Kando}, \bibinfo{person}{Evangelos Kanoulas}, \bibinfo{person}{Martin Potthast}, \bibinfo{person}{Benno Stein}, {et~al\mbox{.}}} \bibinfo{year}{2023}\natexlab{}.
\newblock \showarticletitle{Perspectives on large language models for relevance judgment}. In \bibinfo{booktitle}{\emph{Proceedings of the 2023 ACM SIGIR International Conference on Theory of Information Retrieval}}. \bibinfo{pages}{39--50}.
\newblock


\bibitem[Farzi and Dietz(2024a)]%
        {farzi2024best}
\bibfield{author}{\bibinfo{person}{Naghmeh Farzi} {and} \bibinfo{person}{Laura Dietz}.} \bibinfo{year}{2024}\natexlab{a}.
\newblock \showarticletitle{Best in Tau@ LLMJudge: Criteria-Based Relevance Evaluation with Llama3}.
\newblock \bibinfo{journal}{\emph{arXiv preprint arXiv:2410.14044}} (\bibinfo{year}{2024}).
\newblock


\bibitem[Farzi and Dietz(2024b)]%
        {farzi2024rubric}
\bibfield{author}{\bibinfo{person}{Naghmeh Farzi} {and} \bibinfo{person}{Laura Dietz}.} \bibinfo{year}{2024}\natexlab{b}.
\newblock \showarticletitle{Pencils Down! Automatic Rubric-based Evaluation of Retrieve/Generate Systems}. In \bibinfo{booktitle}{\emph{Proceedings of the 2024 ACM SIGIR International Conference on Theory of Information Retrieval}} (Washington DC, USA) \emph{(\bibinfo{series}{ICTIR '24})}. \bibinfo{publisher}{Association for Computing Machinery}, \bibinfo{address}{New York, NY, USA}, \bibinfo{pages}{175–184}.
\newblock
\showISBNx{9798400706813}
\urldef\tempurl%
\url{https://doi.org/10.1145/3664190.3672511}
\showDOI{\tempurl}


\bibitem[Harman(1992)]%
        {DBLP:conf/trec/1992}
\bibfield{editor}{\bibinfo{person}{Donna~K. Harman}} (Ed.). \bibinfo{year}{1992}\natexlab{}.
\newblock \bibinfo{booktitle}{\emph{Proceedings of The First Text REtrieval Conference, {TREC} 1992, Gaithersburg, Maryland, USA, November 4-6, 1992}}. \bibinfo{series}{{NIST} Special Publication}, Vol.~\bibinfo{volume}{500-207}. \bibinfo{publisher}{National Institute of Standards and Technology {(NIST)}}.
\newblock
\urldef\tempurl%
\url{http://trec.nist.gov/pubs/trec1/t1\_proceedings.html}
\showURL{%
\tempurl}


\bibitem[Joachims et~al\mbox{.}(2017)]%
        {JoachimsSwaminathanSchnabel2017}
\bibfield{author}{\bibinfo{person}{Thorsten Joachims}, \bibinfo{person}{Adith Swaminathan}, {and} \bibinfo{person}{Tobias Schnabel}.} \bibinfo{year}{2017}\natexlab{}.
\newblock \showarticletitle{Unbiased Learning-to-Rank with Biased Feedback}. In \bibinfo{booktitle}{\emph{Proceedings of the Tenth {ACM} International Conference on Web Search and Data Mining, {WSDM} 2017, Cambridge, United Kingdom, February 6-10, 2017}}. \bibinfo{publisher}{{ACM}}, \bibinfo{pages}{781--789}.
\newblock
\urldef\tempurl%
\url{https://doi.org/10.1145/3018661.3018699}
\showDOI{\tempurl}


\bibitem[Jones(1995)]%
        {DBLP:journals/ipm/Jones95}
\bibfield{author}{\bibinfo{person}{Karen~Sparck Jones}.} \bibinfo{year}{1995}\natexlab{}.
\newblock \showarticletitle{Reflections on {TREC}}.
\newblock \bibinfo{journal}{\emph{Inf. Process. Manag.}} \bibinfo{volume}{31}, \bibinfo{number}{3} (\bibinfo{year}{1995}), \bibinfo{pages}{291--314}.
\newblock
\urldef\tempurl%
\url{https://doi.org/10.1016/0306-4573(94)00048-8}
\showDOI{\tempurl}


\bibitem[Kando et~al\mbox{.}(1999)]%
        {kando1999overview}
\bibfield{author}{\bibinfo{person}{Noriko Kando}, \bibinfo{person}{Kazuko Kuriyama}, \bibinfo{person}{Toshihiko Nozue}, \bibinfo{person}{Koji Eguchi}, \bibinfo{person}{Hiroyuki Kato}, {and} \bibinfo{person}{Souichiro Hidaka}.} \bibinfo{year}{1999}\natexlab{}.
\newblock \showarticletitle{Overview of IR tasks at the first NTCIR workshop}. In \bibinfo{booktitle}{\emph{Proceedings of the first NTCIR workshop on research in Japanese text retrieval and term recognition}}. \bibinfo{pages}{11--44}.
\newblock


\bibitem[Liang et~al\mbox{.}(2022)]%
        {liang2022holistic}
\bibfield{author}{\bibinfo{person}{Percy Liang}, \bibinfo{person}{Rishi Bommasani}, \bibinfo{person}{Tony Lee}, \bibinfo{person}{Dimitris Tsipras}, \bibinfo{person}{Dilara Soylu}, \bibinfo{person}{Michihiro Yasunaga}, \bibinfo{person}{Yian Zhang}, \bibinfo{person}{Deepak Narayanan}, \bibinfo{person}{Yuhuai Wu}, \bibinfo{person}{Ananya Kumar}, {et~al\mbox{.}}} \bibinfo{year}{2022}\natexlab{}.
\newblock \showarticletitle{Holistic evaluation of language models}.
\newblock \bibinfo{journal}{\emph{arXiv preprint arXiv:2211.09110}} (\bibinfo{year}{2022}).
\newblock


\bibitem[MacAvaney and Soldaini(2023)]%
        {DBLP:conf/sigir/MacAvaneyS23}
\bibfield{author}{\bibinfo{person}{Sean MacAvaney} {and} \bibinfo{person}{Luca Soldaini}.} \bibinfo{year}{2023}\natexlab{}.
\newblock \showarticletitle{One-Shot Labeling for Automatic Relevance Estimation}. In \bibinfo{booktitle}{\emph{Proceedings of the 46th International {ACM} {SIGIR} Conference on Research and Development in Information Retrieval, {SIGIR} 2023, Taipei, Taiwan, July 23-27, 2023}}. \bibinfo{publisher}{{ACM}}, \bibinfo{pages}{2230--2235}.
\newblock
\urldef\tempurl%
\url{https://doi.org/10.1145/3539618.3592032}
\showDOI{\tempurl}


\bibitem[Meng et~al\mbox{.}(2024)]%
        {meng2024query}
\bibfield{author}{\bibinfo{person}{Chuan Meng}, \bibinfo{person}{Negar Arabzadeh}, \bibinfo{person}{Arian Askari}, \bibinfo{person}{Mohammad Aliannejadi}, {and} \bibinfo{person}{Maarten de Rijke}.} \bibinfo{year}{2024}\natexlab{}.
\newblock \showarticletitle{Query performance prediction using relevance judgments generated by large language models}.
\newblock \bibinfo{journal}{\emph{arXiv preprint arXiv:2404.01012}} (\bibinfo{year}{2024}).
\newblock


\bibitem[Nguyen et~al\mbox{.}(2016)]%
        {DBLP:conf/nips/NguyenRSGTMD16}
\bibfield{author}{\bibinfo{person}{Tri Nguyen}, \bibinfo{person}{Mir Rosenberg}, \bibinfo{person}{Xia Song}, \bibinfo{person}{Jianfeng Gao}, \bibinfo{person}{Saurabh Tiwary}, \bibinfo{person}{Rangan Majumder}, {and} \bibinfo{person}{Li Deng}.} \bibinfo{year}{2016}\natexlab{}.
\newblock \showarticletitle{{MS} {MARCO:} {A} Human Generated MAchine Reading COmprehension Dataset}. In \bibinfo{booktitle}{\emph{Proceedings of the Workshop on Cognitive Computation: Integrating neural and symbolic approaches 2016 co-located with the 30th Annual Conference on Neural Information Processing Systems {(NIPS} 2016), Barcelona, Spain, December 9, 2016}} \emph{(\bibinfo{series}{{CEUR} Workshop Proceedings}, Vol.~\bibinfo{volume}{1773})}. \bibinfo{publisher}{CEUR-WS.org}.
\newblock
\urldef\tempurl%
\url{https://ceur-ws.org/Vol-1773/CoCoNIPS\_2016\_paper9.pdf}
\showURL{%
\tempurl}


\bibitem[Parry et~al\mbox{.}(2024)]%
        {DBLP:conf/ecir/ParryFMPH24}
\bibfield{author}{\bibinfo{person}{Andrew Parry}, \bibinfo{person}{Maik Fr{\"{o}}be}, \bibinfo{person}{Sean MacAvaney}, \bibinfo{person}{Martin Potthast}, {and} \bibinfo{person}{Matthias Hagen}.} \bibinfo{year}{2024}\natexlab{}.
\newblock \showarticletitle{Analyzing Adversarial Attacks on Sequence-to-Sequence Relevance Models}. In \bibinfo{booktitle}{\emph{Advances in Information Retrieval - 46th European Conference on Information Retrieval, {ECIR} 2024, Glasgow, UK, March 24-28, 2024, Proceedings, Part {II}}} \emph{(\bibinfo{series}{Lecture Notes in Computer Science}, Vol.~\bibinfo{volume}{14609})}. \bibinfo{publisher}{Springer}, \bibinfo{pages}{286--302}.
\newblock
\urldef\tempurl%
\url{https://doi.org/10.1007/978-3-031-56060-6\_19}
\showDOI{\tempurl}


\bibitem[Peters(2001)]%
        {DBLP:conf/clef/2000}
\bibfield{editor}{\bibinfo{person}{Carol Peters}} (Ed.). \bibinfo{year}{2001}\natexlab{}.
\newblock \bibinfo{booktitle}{\emph{Cross-Language Information Retrieval and Evaluation, Workshop of Cross-Language Evaluation Forum, {CLEF} 2000, Lisbon, Portugal, September 21-22, 2000, Revised Papers}}. \bibinfo{series}{Lecture Notes in Computer Science}, Vol.~\bibinfo{volume}{2069}. \bibinfo{publisher}{Springer}.
\newblock
\showISBNx{3-540-42446-6}
\urldef\tempurl%
\url{https://doi.org/10.1007/3-540-44645-1}
\showDOI{\tempurl}


\bibitem[Rahmani et~al\mbox{.}(2024a)]%
        {rahmani2024synthetic}
\bibfield{author}{\bibinfo{person}{Hossein~A Rahmani}, \bibinfo{person}{Nick Craswell}, \bibinfo{person}{Emine Yilmaz}, \bibinfo{person}{Bhaskar Mitra}, {and} \bibinfo{person}{Daniel Campos}.} \bibinfo{year}{2024}\natexlab{a}.
\newblock \showarticletitle{Synthetic Test Collections for Retrieval Evaluation}.
\newblock \bibinfo{journal}{\emph{arXiv preprint arXiv:2405.07767}} (\bibinfo{year}{2024}).
\newblock


\bibitem[Rahmani et~al\mbox{.}(2024b)]%
        {rahmani2024report}
\bibfield{author}{\bibinfo{person}{Hossein~A Rahmani}, \bibinfo{person}{Clemencia Siro}, \bibinfo{person}{Mohammad Aliannejadi}, \bibinfo{person}{Nick Craswell}, \bibinfo{person}{Charles~LA Clarke}, \bibinfo{person}{Guglielmo Faggioli}, \bibinfo{person}{Bhaskar Mitra}, \bibinfo{person}{Paul Thomas}, {and} \bibinfo{person}{Emine Yilmaz}.} \bibinfo{year}{2024}\natexlab{b}.
\newblock \showarticletitle{Report on the 1st workshop on large language model for evaluation in information retrieval (llm4eval 2024) at sigir 2024}.
\newblock \bibinfo{journal}{\emph{arXiv preprint arXiv:2408.05388}} (\bibinfo{year}{2024}).
\newblock


\bibitem[Rahmani et~al\mbox{.}(2024c)]%
        {rahmani2024llm4eval}
\bibfield{author}{\bibinfo{person}{Hossein~A. Rahmani}, \bibinfo{person}{Clemencia Siro}, \bibinfo{person}{Mohammad Aliannejadi}, \bibinfo{person}{Nick Craswell}, \bibinfo{person}{Charles L.~A. Clarke}, \bibinfo{person}{Guglielmo Faggioli}, \bibinfo{person}{Bhaskar Mitra}, \bibinfo{person}{Paul Thomas}, {and} \bibinfo{person}{Emine Yilmaz}.} \bibinfo{year}{2024}\natexlab{c}.
\newblock \showarticletitle{LLM4Eval: Large Language Model for Evaluation in IR}. In \bibinfo{booktitle}{\emph{Proceedings of the 47th International ACM SIGIR Conference on Research and Development in Information Retrieval}} (Washington DC, USA) \emph{(\bibinfo{series}{SIGIR '24})}. \bibinfo{publisher}{Association for Computing Machinery}, \bibinfo{address}{New York, NY, USA}, \bibinfo{pages}{3040–3043}.
\newblock
\showISBNx{9798400704314}
\urldef\tempurl%
\url{https://doi.org/10.1145/3626772.3657992}
\showDOI{\tempurl}


\bibitem[Rahmani et~al\mbox{.}(2024d)]%
        {rahmani2024judgeblender}
\bibfield{author}{\bibinfo{person}{Hossein~A Rahmani}, \bibinfo{person}{Emine Yilmaz}, \bibinfo{person}{Nick Craswell}, {and} \bibinfo{person}{Bhaskar Mitra}.} \bibinfo{year}{2024}\natexlab{d}.
\newblock \showarticletitle{JudgeBlender: Ensembling Judgments for Automatic Relevance Assessment}.
\newblock \bibinfo{journal}{\emph{arXiv preprint arXiv:2412.13268}} (\bibinfo{year}{2024}).
\newblock


\bibitem[Rahmani et~al\mbox{.}(2024e)]%
        {rahmani2024llmjudge}
\bibfield{author}{\bibinfo{person}{Hossein~A Rahmani}, \bibinfo{person}{Emine Yilmaz}, \bibinfo{person}{Nick Craswell}, \bibinfo{person}{Bhaskar Mitra}, \bibinfo{person}{Paul Thomas}, \bibinfo{person}{Charles~LA Clarke}, \bibinfo{person}{Mohammad Aliannejadi}, \bibinfo{person}{Clemencia Siro}, {and} \bibinfo{person}{Guglielmo Faggioli}.} \bibinfo{year}{2024}\natexlab{e}.
\newblock \showarticletitle{Llmjudge: Llms for relevance judgments}.
\newblock \bibinfo{journal}{\emph{arXiv preprint arXiv:2408.08896}} (\bibinfo{year}{2024}).
\newblock


\bibitem[Sanderson(2010)]%
        {DBLP:journals/ftir/Sanderson10}
\bibfield{author}{\bibinfo{person}{Mark Sanderson}.} \bibinfo{year}{2010}\natexlab{}.
\newblock \showarticletitle{Test Collection Based Evaluation of Information Retrieval Systems}.
\newblock \bibinfo{journal}{\emph{Found. Trends Inf. Retr.}} \bibinfo{volume}{4}, \bibinfo{number}{4} (\bibinfo{year}{2010}), \bibinfo{pages}{247--375}.
\newblock
\urldef\tempurl%
\url{https://doi.org/10.1561/1500000009}
\showDOI{\tempurl}


\bibitem[Scells et~al\mbox{.}(2022)]%
        {DBLP:conf/sigir/ScellsZZ22}
\bibfield{author}{\bibinfo{person}{Harrisen Scells}, \bibinfo{person}{Shengyao Zhuang}, {and} \bibinfo{person}{Guido Zuccon}.} \bibinfo{year}{2022}\natexlab{}.
\newblock \showarticletitle{Reduce, Reuse, Recycle: Green Information Retrieval Research}. In \bibinfo{booktitle}{\emph{{SIGIR} '22: The 45th International {ACM} {SIGIR} Conference on Research and Development in Information Retrieval, Madrid, Spain, July 11 - 15, 2022}}. \bibinfo{publisher}{{ACM}}, \bibinfo{pages}{2825--2837}.
\newblock
\urldef\tempurl%
\url{https://doi.org/10.1145/3477495.3531766}
\showDOI{\tempurl}


\bibitem[Soboroff(2024)]%
        {DBLP:journals/corr/abs-2409-15133}
\bibfield{author}{\bibinfo{person}{Ian Soboroff}.} \bibinfo{year}{2024}\natexlab{}.
\newblock \showarticletitle{Don't Use LLMs to Make Relevance Judgments}.
\newblock \bibinfo{journal}{\emph{CoRR}}  \bibinfo{volume}{abs/2409.15133} (\bibinfo{year}{2024}).
\newblock
\urldef\tempurl%
\url{https://doi.org/10.48550/ARXIV.2409.15133}
\showDOI{\tempurl}
\showeprint[arXiv]{2409.15133}


\bibitem[Sun et~al\mbox{.}(2023)]%
        {Sun2023IsCG}
\bibfield{author}{\bibinfo{person}{Weiwei Sun}, \bibinfo{person}{Lingyong Yan}, \bibinfo{person}{Xinyu Ma}, \bibinfo{person}{Pengjie Ren}, \bibinfo{person}{Dawei Yin}, {and} \bibinfo{person}{Zhaochun Ren}.} \bibinfo{year}{2023}\natexlab{}.
\newblock \showarticletitle{Is ChatGPT Good at Search? Investigating Large Language Models as Re-Ranking Agent}.
\newblock \bibinfo{journal}{\emph{ArXiv}}  \bibinfo{volume}{abs/2304.09542} (\bibinfo{year}{2023}).
\newblock


\bibitem[Thomas et~al\mbox{.}(2023)]%
        {thomas2023large}
\bibfield{author}{\bibinfo{person}{Paul Thomas}, \bibinfo{person}{Seth Spielman}, \bibinfo{person}{Nick Craswell}, {and} \bibinfo{person}{Bhaskar Mitra}.} \bibinfo{year}{2023}\natexlab{}.
\newblock \showarticletitle{Large language models can accurately predict searcher preferences}.
\newblock \bibinfo{journal}{\emph{arXiv preprint arXiv:2309.10621}} (\bibinfo{year}{2023}).
\newblock


\bibitem[Upadhyay et~al\mbox{.}(2024)]%
        {upadhyay2024umbrela}
\bibfield{author}{\bibinfo{person}{Shivani Upadhyay}, \bibinfo{person}{Ronak Pradeep}, \bibinfo{person}{Nandan Thakur}, \bibinfo{person}{Nick Craswell}, {and} \bibinfo{person}{Jimmy Lin}.} \bibinfo{year}{2024}\natexlab{}.
\newblock \showarticletitle{UMBRELA: UMbrela is the (Open-Source Reproduction of the) Bing RELevance Assessor}.
\newblock \bibinfo{journal}{\emph{arXiv preprint arXiv:2406.06519}} (\bibinfo{year}{2024}).
\newblock


\bibitem[Vaswani et~al\mbox{.}(2017)]%
        {DBLP:conf/nips/VaswaniSPUJGKP17}
\bibfield{author}{\bibinfo{person}{Ashish Vaswani}, \bibinfo{person}{Noam Shazeer}, \bibinfo{person}{Niki Parmar}, \bibinfo{person}{Jakob Uszkoreit}, \bibinfo{person}{Llion Jones}, \bibinfo{person}{Aidan~N. Gomez}, \bibinfo{person}{Lukasz Kaiser}, {and} \bibinfo{person}{Illia Polosukhin}.} \bibinfo{year}{2017}\natexlab{}.
\newblock \showarticletitle{Attention is All you Need}. In \bibinfo{booktitle}{\emph{Advances in Neural Information Processing Systems 30: Annual Conference on Neural Information Processing Systems 2017, December 4-9, 2017, Long Beach, CA, {USA}}}. \bibinfo{pages}{5998--6008}.
\newblock
\urldef\tempurl%
\url{https://proceedings.neurips.cc/paper/2017/hash/3f5ee243547dee91fbd053c1c4a845aa-Abstract.html}
\showURL{%
\tempurl}


\bibitem[Wang et~al\mbox{.}(2024)]%
        {DBLP:journals/corr/abs-2408-09713}
\bibfield{author}{\bibinfo{person}{Haijin Wang}, \bibinfo{person}{Mianrong Zhang}, \bibinfo{person}{Zheng Chen}, \bibinfo{person}{Nan Shang}, \bibinfo{person}{Shangheng Yao}, \bibinfo{person}{Fushuan Wen}, {and} \bibinfo{person}{Junhua Zhao}.} \bibinfo{year}{2024}\natexlab{}.
\newblock \showarticletitle{Carbon Footprint Accounting Driven by Large Language Models and Retrieval-augmented Generation}.
\newblock \bibinfo{journal}{\emph{CoRR}}  \bibinfo{volume}{abs/2408.09713} (\bibinfo{year}{2024}).
\newblock
\urldef\tempurl%
\url{https://doi.org/10.48550/ARXIV.2408.09713}
\showDOI{\tempurl}
\showeprint[arXiv]{2408.09713}


\bibitem[Zuccon et~al\mbox{.}(2023)]%
        {DBLP:conf/ictir/ZucconSZ23}
\bibfield{author}{\bibinfo{person}{Guido Zuccon}, \bibinfo{person}{Harrisen Scells}, {and} \bibinfo{person}{Shengyao Zhuang}.} \bibinfo{year}{2023}\natexlab{}.
\newblock \showarticletitle{Beyond {CO2} Emissions: The Overlooked Impact of Water Consumption of Information Retrieval Models}. In \bibinfo{booktitle}{\emph{Proceedings of the 2023 {ACM} {SIGIR} International Conference on Theory of Information Retrieval, {ICTIR} 2023, Taipei, Taiwan, 23 July 2023}}. \bibinfo{publisher}{{ACM}}, \bibinfo{pages}{283--289}.
\newblock
\urldef\tempurl%
\url{https://doi.org/10.1145/3578337.3605121}
\showDOI{\tempurl}


\end{thebibliography}
